\documentclass[5p,times]{elsarticle}
\pdfoutput=1
\usepackage{lineno,hyperref}
\usepackage{amssymb}
\usepackage{amsmath}
\journal{Computing in Astronomy}

\usepackage{listings}
\usepackage{color}
\definecolor{mygreen}{rgb}{0.3,0.5,0}
\definecolor{myred}{rgb}{0.8,0,0}
\definecolor{mygray}{rgb}{0.5,0.5,0.5}

\lstdefinestyle{qasession}{
  language=XML,
  frame=tb,
  aboveskip=1mm,
  belowskip=1mm,
  columns=flexible,
  numbers=left,
  firstnumber=1,
  numberstyle={\tiny\color{mygray}},
  basicstyle={\footnotesize\ttfamily},
  moredelim=[s][\color{mygreen}]{?}{:},
  escapechar={*}
}

\lstdefinestyle{skifile}{
  language=XML,
  frame=tb,
  aboveskip=1mm,
  belowskip=1mm,
  columns=flexible,
  numbers=left,
  firstnumber=1,
  numberstyle={\tiny\color{mygray}},
  basicstyle={\footnotesize\ttfamily},
  stringstyle=\color{myred},
  keywordstyle=\color{mygreen},
  tagstyle=\color{black},
  usekeywordsintag=true,
  markfirstintag=true,
  showstringspaces=false,
}

\lstdefinestyle{skitex}{
  language=XML,
  frame=tb,
  aboveskip=1mm,
  belowskip=1mm,
  columns=flexible,
  numbers=left,
  firstnumber=1,
  numberstyle={\tiny\color{mygray}},
  basicstyle={\small},
  escapechar={*}
}

\lstdefinestyle{cpp}{
  language=C++,
  frame=tb,
  aboveskip=1mm,
  belowskip=1mm,
  columns=flexible,
  numbers=left,
  firstnumber=1,
  numberstyle={\tiny\color{mygray}},
  basicstyle={\footnotesize\ttfamily},
  stringstyle=\color{myred},
  keywordstyle=\color{mygreen},
  showstringspaces=false,
}

\newcommand{\cpp}[1]{{\small\ttfamily #1}}
\newcommand{\role}[1]{{\small\itshape\sffamily #1}}

\bibpunct{(}{)}{;}{a}{}{,} 
\begin{document}
\begin{frontmatter}
\title{SKIRT: an Advanced Dust Radiative Transfer Code with a User-Friendly Architecture}
\author{Peter Camps\corref{correspondingauthor}}
\ead{peter.camps@ugent.be}
\cortext[correspondingauthor]{Corresponding author}
\author{Maarten Baes\corref{none}}
\address{Sterrenkundig Observatorium, Universiteit Gent, Krijgslaan 281, B-9000 Gent, Belgium}

\begin{abstract}
We discuss the architecture and design principles that underpin the latest version of SKIRT, a state-of-the-art open source code for simulating continuum radiation transfer in dusty astrophysical systems, such as spiral galaxies and accretion disks. SKIRT employs the Monte Carlo technique to emulate the relevant physical processes including scattering, absorption and emission by the dust. The code features a wealth of built-in geometries, radiation source spectra, dust characterizations, dust grids, and detectors, in addition to various mechanisms for importing snapshots generated by hydrodynamical simulations. The configuration for a particular simulation is defined at run-time through a user-friendly interface suitable for both occasional and power users. These capabilities are enabled by careful C++ code design. The programming interfaces between components are well defined and narrow. Adding a new feature is usually as simple as adding another class; the user interface automatically adjusts to allow configuring the new options. We argue that many scientific codes, like SKIRT, can benefit from careful object-oriented design and from a friendly user interface, even if it is not a graphical user interface.
\end{abstract}

\end{frontmatter}

\section{Introduction}

The presence of even a small fraction of dust can have a substantial impact on the radiation traversing and exiting an astrophysical system. In a typical spiral galaxy viewed edge-on, for example, the central dust lane blocks most of the starlight in the UV and optical wavelength range and re-emits the absorbed energy in the infrared and sub-millimeter regime \citep[e.g.\@][]{2013A&A...556A..54V}. Simulating the precise effect of the dust is not trivial. Aniso\-tropic scattering by the dust couples all lines of sight, and dust absorption/emission couples all wavelengths. As a result, the radiative transfer equation is highly nonlocal and nonlinear \citep{2013ARA&A..51...63S}. In most astrophysical systems, at least part of the dust is not in local thermal equilibrium with the radiation field, complicating the calculations even more \citep{2001ApJ...551..807D}. And finally, realistic scenarios involve complex 3D geometries such as spiral arms, randomly placed clumps, or snapshots taken from a hydrodynamical simulation.

Because of these complexities, most dust radiative transfer codes use the Monte Carlo technique to tackle the problem; see e.g. the reviews by \citet{2011BASI...39..101W} and \citet{2013ARA&A..51...63S}. The radiation field is represented as a stream of discrete photon packages. A simulation follows the individual path of each photon package through the dusty medium. The trajectory is governed by various events determined statistically by drawing random numbers from the appropriate probability distribution. Typically, a photon package is emitted, undergoes a number of scattering events, and is finally either absorbed or leaves the system. The Monte Carlo technique is conceptually simple and allows efficient radiative transfer calculations for complex problems. However, due to the randomization process, the results inherently contain a certain level of Poisson noise.

SKIRT is a state-of-the-art Monte Carlo dust radiative transfer code. It implements the common optimization techniques, such as peel-off at emission and scattering events \citep{1984ApJ...278..186Y}, continuous absorption \citep{1999A&A...344..282L,2003A&A...399..703N}, and forced scattering \citep{Cashwell1959}, and includes novel techniques such as the library mechanism described in \citet{2011ApJS..196...22B}. The code is registered in the Astrophysics Source Code Library with identifier \mbox{\emph{ascl:1109.003}}. Earlier versions were described in \citet{2003MNRAS.343.1081B} and in \citet{2011ApJS..196...22B}. Here we present the latest, substantially revised version of SKIRT, which is fully documented\footnote{SKIRT documentation: http://www.skirt.ugent.be} and publicly available from a GitHub code repository\footnote{SKIRT code repository: https://github.com/skirt/skirt}.

\citet{2012MNRAS.420.2756S} have used SKIRT to investigate the emission of active galactic nuclei (AGN) dusty tori in the infrared domain, modeling the dusty torus as a two-phase medium with high-density clumps and a low-density medium filling the space between the clumps. The resulting SED database has been made public as described in \citet{2012BlgAJ..18c...3S}. As can be expected, SKIRT is often used to solve the \emph{inverse} radiative transfer problem, where the goal is to recover the actual 3D distribution of radiation sources and dust by fitting the results of radiative transfer simulations to observational data. This is a nontrivial task, since the underlying model typically has a large number of free parameters. For example, SKIRT has been used to perform detailed studies of the dust energy balance in the edge-on spiral galaxies UGC4754, NGC 4565, and M104 \citep{2010AA...518L..39B,2012MNRAS.427.2797D,2012MNRAS.419..895D}. These studies found an inconsistency in the dust energy budget, suggesting that a sizable fraction of the total dust reservoir consists of a clumpy distribution with no associated young stellar sources.

Rather than using a manual trial and error procedure, \citet{2013A&A...550A..74D} describe FitSKIRT, a code that automatically fits a 3D model to observed images of a dusty galaxy by matching the output of SKIRT radiative transfer simulations to the data. They apply FitSKIRT to automatically determine the intrinsic distribution of stars and dust in the galaxy NGC\,4013. \citet{2014MNRAS.441..869D} use FitSKIRT to investigate interstellar dust in a sample of 12 edge-on galaxies, simultaneously reproducing the $g$-, $r$-, $i$- and $z$-band observations from a model with 19 free parameters without human intervention.

In this article, we do not discuss the results obtained with SKIRT, nor the Monte Carlo radiative transfer techniques implemented in the code. Instead we focus on the software design choices involved with setting up the simulation model. Many scientific codes require a user to hard-code the model makeup for each distinct problem. In contrast, our strategy with SKIRT in recent years has been to continuously add new features without removing existing capabilities. Consequently, SKIRT now offers a wealth of configurable components that are ready to use without any programming at all, especially in the areas where the code has been most often applied. An ad-hoc approach to including all of this functionality would have lead to source code that is hard to understand, maintain, and use. Instead we developed a modular, generic software architecture that can support the wide range of built-in components and options in SKIRT in a developer- and user-friendly way.

In Sect.\,\ref{sec:features} we first provide an overview of SKIRT's features, including the user interface for configuring a particular simulation. In Sect.\,\ref{sec:architecture} we then discuss the design goals for the latest revision of the code, we describe the overall architecture, and we zoom in on a few key aspects of the design, such as the mechanism that automatically adjusts the user interface to accommodate new features. In Sect.\,\ref{sec:conclusions} we finally argue that many scientific codes, like SKIRT, can benefit from careful object-oriented design and from a friendly user interface, even if it is not a graphical user interface.

\section{Features}
\label{sec:features}

\subsection{Overview}

SKIRT is a Monte Carlo continuum radiative transfer code for simulating the effect of dust on radiation in static astrophysical systems. It is assumed that the radiation traverses the system much faster than the time scale on which the system evolves. SKIRT offers full treatment of absorption and multiple an\-isotropic scattering by the dust, computes the temperature distribution of the dust and the thermal dust re-emission self-con\-sistently, and supports stochastic heating of small grains using an efficient library approach. It handles multiple dust mixtures and arbitrary 3D geometries for radiation sources and dust components, and offers a variety of simulated instruments for measuring the radiation field from any angle. The code operates efficiently and is parallelized on shared memory systems.

SKIRT is a console application and is completely written in C++. It can easily be deployed on any Unix system, including for example Ubuntu and Mac OS X. The code has no compile-time options; all built-in components and capabilities are configured at run-time. The first-time user would use the interactive query and answer mechanism (in a terminal window) to configure a particular simulation. The smart mechanism guides the user through all possible options, narrowing down the possibilities based on earlier choices. The complete configuration for the simulation is then saved as a SKIRT parameter file in XML (eXtensible Markup Language) format, which can be easily viewed and adjusted in a regular text editor, even by an occasional user.

\subsection{Configuring a simulation}

\begin{figure}
  \centering
  \includegraphics[width=0.70\columnwidth]{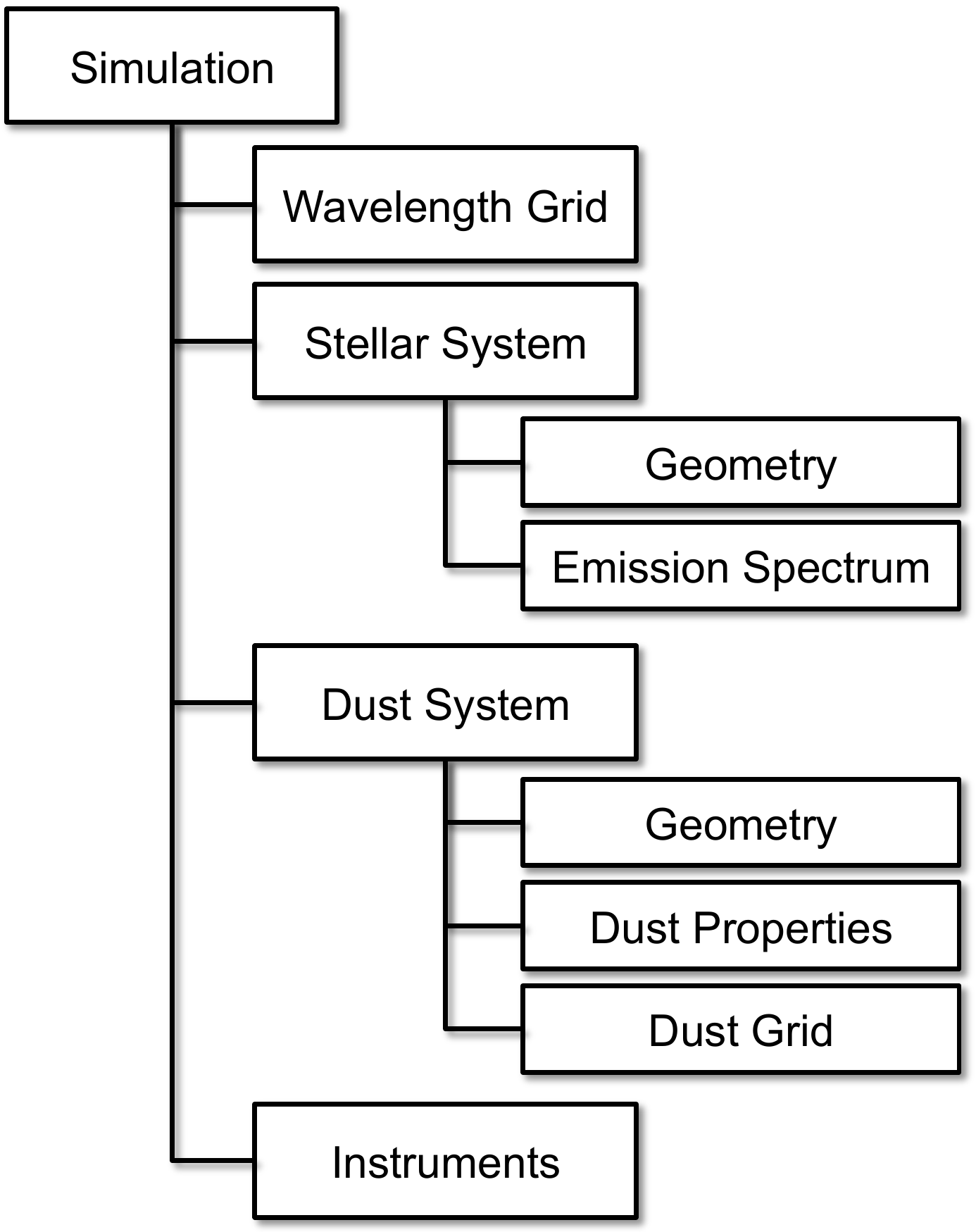}
  \caption{A schematic representation of the items to be configured for a particular SKIRT simulation.}
  \label{fig:configuration}
\end{figure}

Figure\,\ref{fig:configuration} illustrates the structure of a SKIRT simulation. Each of the building blocks offers several alternatives and options that can be configured in the parameter file. At the top level, for example, SKIRT supports two simulation types: oligochromatic and panchromatic. An \emph{oligochromatic} simulation operates at just one or a small number of distinct wavelengths. It handles absorption and scattering by dust grains, but it doesn't support thermal dust emission. There is no way to compute the dust temperature without integrating the absorbed radiation energy over an appropriate wavelength range. This basic simulation mode is appropriate for studying optical wavelengths, since the dust emission is negligible there. A \emph{panchromatic} simulation operates over a broad range of wavelengths. These simulations can handle thermal dust emission as well as absorption and scattering, and thus many more options need to be configured.

The wavelength grid for an oligochromatic simulation is simply a short list of distinct wavelengths. A panchromatic simulation employs a grid over a range that typically extends from UV to millimeter wavelengths. SKIRT offers a plain logarithmic wavelength grid and a nested logarithmic wavelength grid, providing a higher resolution in some subset of the range. The user can specify the wavelength range and the number of grid points. Alternatively SKIRT can read a custom wavelength grid from a text data file that lists the grid points.

\begin{table*}
  \caption{An overview of built-in components that can be used for defining (a) the spatial distribution of radiation sources and dust components, (b) the spectral energy distribution of radiation sources, (c) the properties of the dust mixture, and (d) the spatial discretization of the dust medium.}
  \label{table:components}
  \centering
  \footnotesize
  \begin{tabular}{lll}
    \\ \hline
    (a) \textbf{Geometries} 
      & \emph{Spherically symmetric} \\ 
        & \quad PointGeometry & single point \\
        & \quad PlummerGeometry & classical Plummer sphere \citep{1911MNRAS..71..460P,1987MNRAS.224...13D} \\
        & \quad SersicGeometry & spherical model with a S\'{e}rsic surface brightness profile \citep{1963BAAA....6...41S,1999AA...352..447C} \\
        & \quad EinastoGeometry & spherical model with an Einasto density profile \citep{1965TrAlm...5...87E,2012AA...540A..70R} \\
        & \quad GammaGeometry & spherical model with a gamma density profile \citep{1993MNRAS.265..250D,1994AJ....107..634T} \\
        & \quad ShellGeometry & spherical shell where the density behaves as a power law between an inner and an outer radius \\
      & \emph{Axisymmetric} \\ 
        & \quad ExpDiskGeometry & optionally truncated exponential profile in both radial and vertical directions \citep{1986AA...157..230V} \\
        & \quad RingGeometry & ring with gaussian profile in the radial direction and exponential fall-off in the vertical direction \\
        & \quad TorusGeometry & torus with radial power-law profile within opening angle \citep{2012MNRAS.420.2756S,1994MNRAS.268..235G} \\
        & \quad GaussianGeometry & model with gaussian distribution in the radial and the vertical direction \\
        & \quad MGEGeometry & geometry defined by a Multi-Gaussian Expansion \citep{1994AA...285..723E,2002MNRAS.333..400C} \\
      & \emph{No symmetries} \\ 
        & \quad ExpDiskSpiralArmsG... & double-exponential profile with a spiral arm perturbation \citep{2000AA...353..117M} \\
        & \quad AdaptiveMeshGeometry & density distribution defined over an adaptive mesh grid, imported from a data file \\
        & \quad VoronoiGeometry & density distribution defined over a Voronoi tessellation, imported from a data file \\
      & \emph{Decorators} \\ 
        & \quad OffsetGeometry & applies an arbitrary offset to any other geometry \\
        & \quad ClumpyGeometry & replaces a portion of the mass in any geometry by randomly placed clumps \\
        & \quad SpheroidalGeometry & transforms any geometry to a spheroidal counterpart \\
        & \quad TriaxialGeometry & transforms any geometry to a triaxial counterpart \\
      & \emph{Anisotropic} \\ 
        & \quad NetzerGeometry & point source with the anisotropic radiation profile of an accretion disk \citep{1987MNRAS.225...55N} \\
    (b) \textbf{SEDs}
      & \emph{Simple} \\ 
        & \quad BlackBodySED & classical black body spectrum for a given temperature \\
        & \quad PegaseSED & SED templates for elliptical, lenticular and spiral galaxies \citep{1997AA...326..950F} \\
        & \quad QuasarSED & SED template for a quasar \citep{2005AA...437..861S} \\
        & \quad StarburstSED & SED templates for a starbursting stellar population with given metallicity \citep{1999ApJS..123....3L} \\
        & \quad SunSED & the solar spectrum \\
        & \quad FileSED & arbitrary spectrum imported from a data file \\
      & \emph{Families} \\ 
        & \quad BruzualCharlotSED & stellar population SEDs parameterized on metallicity and age \citep{2003MNRAS.344.1000B} \\
        & \quad MarastonSED & stellar population SEDs parameterized on metallicity and age \citep{1998MNRAS.300..872M} \\
        & \quad KuruczSED & stellar SEDs parameterized on metallicity, effective temperature and surface gravity \citep{1993yCat.6039....0K} \\
        & \quad MappingsSED & starbursting region SEDs par. on metallicity, compactness, pressure and covering factor \citep{2008ApJS..176..438G} \\
    (c) \textbf{Dust Mixes}
      & \emph{Turn-key dust mixes} \\ 
        & \quad DraineLiDustMix & mixture of graphite, silicate and PAH grains \citep{2007ApJ...657..810D} \\
        & \quad MRNDustMix & mixture of graphite and silicate grains \citep{1977ApJ...217..425M,2001ApJ...548..296W} \\
        & \quad WeingartnerDustMix & mixture of graphite, silicate and PAH grains \citep{2001ApJ...548..296W} \\
        & \quad ZubkoDustMix & mixture of graphite, silicate and PAH grains \citep{2004ApJS..152..211Z} \\
      & \emph{Custom dust mixes} \\ 
        & \quad ConfigurableDustMix & custom-configured dust mix given a list of grain compositions and grain size distributions \\
      & \emph{Grain compositions} \\ 
        & \quad DraineGrainComp & optical and calorimetric properties for graphite, silicate and PAH grains \citep{2007ApJ...657..810D} \\
        & \quad DustEmGrainComp & any of the dust grain properties provided with the DustEM code \citep{2011AA...525A.103C} \\
        & \quad ForsteriteGrainComp & Forsterite crystalline silicate grain properties \citep{2001AA...378..228F,2005AA...432..909M,2006MNRAS.370.1599S} \\
        & \quad EnstatiteGrainComp & Enstatite crystalline silicate grain properties \citep{1998AA...339..904J,2005AA...432..909M} \\
      & \emph{Grain size distributions} \\ 
        & \quad PowerLawGrainSize & modified power-law grain size distribution with a form inspired by \citet{2011AA...525A.103C} \\
        & \quad LogNormalGrainSize & modified log-normal grain size distribution with a form inspired by \citet{2011AA...525A.103C} \\
    (d) \textbf{Dust Grids}
      & \emph{Spherically symmetric} \\ 
        & \quad LinSpheDustGrid & spherical grid with regularly spaced cells (radii of cell boundaries are equidistant) \\
        & \quad LogSpheDustGrid & spherical grid with logarithmically spaced cells (central cells have smaller widths) \\
        & \quad PowSpheDustGrid & spherical grid with cells spaced according to a power-law (central cells have smaller widths) \\
      & \emph{Axisymmetric} \\ 
        & \quad LinAxDustGrid & cylindrical grid with regularly spaced cells in both radial and vertical directions \\
        & \quad LogLinAxDustGrid & cylindrical grid with logarithmically spaced cells radially, and regularly spaced cells vertically \\
        & \quad PowAxDustGrid & cylindrical grid where cells are spaced according to a power-law in both directions \\
        & \quad LogPowAxDustGrid & cylindrical grid with logarithmically spaced cells radially, and power-law spaced cells vertically \\
      & \emph{Cuboidial} \\ 
        & \quad LinCubDustGrid & grid with regularly spaced cuboidal cells in the three dimensions \\
        & \quad PowCubDustGrid & grid with cuboidal cells spaced according to a power-law in the three dimensions \\
        & \quad OctTreeDustGrid & octree grid that recursively subdivides cuboidal nodes into eight sub-nodes \citep{2013AA...554A..10S} \\
        & \quad BinTreeDustGrid & $k$-d tree grid that recursively subdivides cuboidal nodes into two sub-nodes \citep{2014AA...561A..77S} \\
      & \emph{Unstructured} \\ 
        & \quad VoronoiDustGrid & unstructured dust grid based on a Voronoi tesselation of 3D space \citep{2013AA...560A..35C} \\
        \hline
  \end{tabular} 
\end{table*}

The spatial distribution of radiation sources and dust is obviously an important part of the simulation model. For this purpose SKIRT offers a number of predefined geometries; the most important ones are listed in Table\,\ref{table:components}(a). Each geometry defines a spatial density distribution, which can be used for radiation sources as well as dust components. Choices include a point-like source and various theoretical models for distributed densities with spherical, cylindrical, or no symmetries. Decorator geometries adjust another geometry by shifting its center to an arbitrary location, deforming a spherical geometry into a spheroidal or triaxial distribution, or adding clumps in random locations. Other geometries can import a density distribution from a data file. Anisotropic radiation sources are supported as well. Multiple geometries can be combined in arbitrary ways, enabling the construction of complex models.

The stellar system describes the radiation sources in the simulation model. For each geometry, the configuration defines the emission spectrum and the luminosity. Table\,\ref{table:components}(b) lists the built-in spectral energy distributions (SEDs), including several well-known parameterized SED families, and the option to import an SED from file. The amount of radiation can be specified through the bolometric luminosity or through the spectral luminosity at the center of a standard wavelength band. SKIRT also includes specialized stellar systems to import a snapshot from a hydrodynamic simulation using smoothed particles (SPH) or an adaptive mesh (AMR). In this case, both the spatial distribution and the emission spectrum in each location are extracted from the input data, for example using a Bruzual-Charlot model based on stellar age and metallicity.

Similarly, the dust system describes the spatial distribution and the properties of the dust in the model. A dust system can have multiple components, each with its own geometry and dust characterization. The amount of dust in each component can be defined simply as a total mass, or by specifying the optical depth along a particular axis. The optical and chemical properties of the dust in each component can be configured in great detail, as described in Sect.\,\ref{sec:dustprops}. Again, there are specialized dust systems to import a snapshot from a hydrodynamic simulation (SPH or AMR). The spatial distribution of the dust is now calculated from the gas density in the input data, assuming that the amount of dust is proportional to the metal fraction in the gas, except in areas where the gas is too hot to form dust.

The dust system also configures the dust grid, i.e.\ the computational structure that is used to discretize the spatial domain under study. The grid partitions the spatial domain in individual dust cells, and all physical variables (dust density, optical properties, radiation field, dust temperature) are considered to be constant in each dust cell. During the radiative transfer simulation, photon packages propagate through the grid and interact with particular cells according to randomly generated events. Since memory requirements and computation time rapidly increase with the number of dust cells, a good grid has smaller cells in areas that require a higher resolution, and larger cells elsewhere. Table\,\ref{table:components}(d) lists the dust grids built into SKIRT. The spherical and cylindrical grids are perfect for simple models with the corresponding symmetries. Linear grids have equidistant grid points; logarithmic and power-law grids place (much) smaller bins in the central areas of the model. Most state-of-the-art simulations use 3D models, however, and SKIRT offers a choice of smart structured and unstructured grids to help optimize accuracy and performance. This is an active field of study in our research group, as described in Sect.\,\ref{sec:dustgrids}.

Finally, the configuration is completed with a number of synthetic instruments, which collect and write down information about the simulated radiation received at some specified viewpoint. The \emph{SED} instrument outputs the spectral energy distribution of the received flux as a text file that can easily be plotted. The \emph{frame} instrument collects a complete 3D data cube (a rectangular frame of flux samples at each simulated wavelength) and outputs the result as a FITS file (Flexible Image Transport System), enabling the use of the standard visualization and data manipulation tools. The instruments can treat the radiation differently depending on its source; for example direct radiation, scattered radiation and dust emission can be recorded separately. The default instruments assume that the distance to the model is very large, so that they can use parallel projection. The perspective instrument, however, can be placed anywhere, even inside the model. It is mostly used to create animations by specifying an instrument per movie frame, with slightly varying position and/or angles.

\subsection{Dust properties}
\label{sec:dustprops}

The dust system in a SKIRT simulation can hold multiple dust components, each with their own spatial distribution and their specific dust characterization. The dust properties applying to a particular dust component are bundled in a building block called a \emph{dust mix}. SKIRT offers several options to configure a dust mix, ranging from very simple to quite involved. Table\,\ref{table:components}(c) lists some of the choices. Each of the turn-key dust mixes implements a particular dust model described in the literature, usually including some specific combination of silicate grains, graphite grains, and polycyclic aromatic hydrocarbon (PAH) molecules, with properties listed in data files and/or approximated by formulae. These dust mixes can be configured simply by supplying their name.

\begin{figure}
  \centering
  \includegraphics[width=0.70\columnwidth]{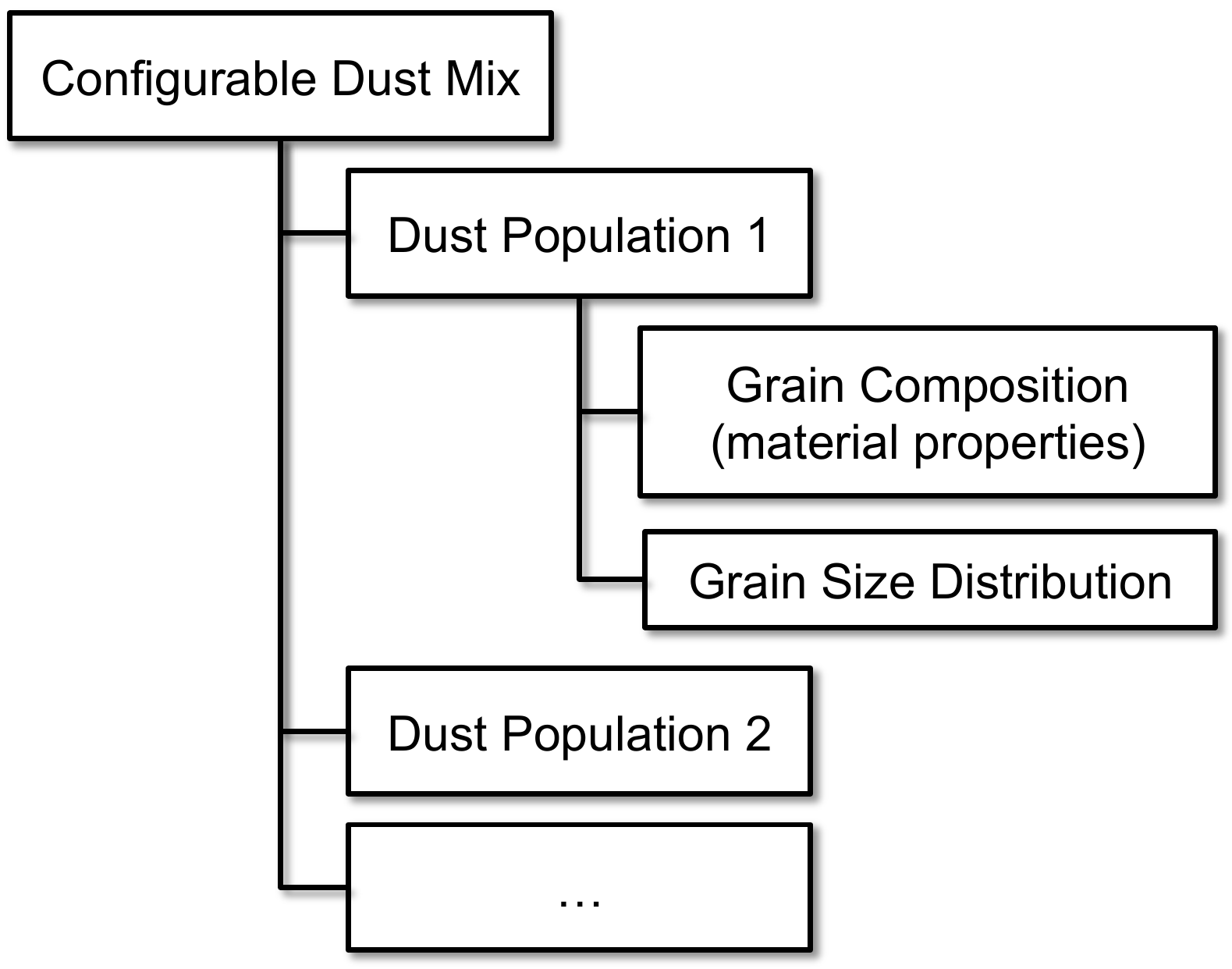}
  \caption{A schematic representation of a dust mix containing multiple dust populations. Each population describes a particular type of grain.}
  \label{fig:dustmix}
\end{figure}

Alternatively, the user can configure a custom dust mix from basic building blocks, as illustrated in Fig.\,\ref{fig:dustmix}. A configurable dust mix holds a distinct dust population for each type of grain material in the mix. For each type of grain material, the dust population specifies the optical and calorimetric material properties and a grain size distribution function. Optical dust properties include the scattering and absorption coefficients $\kappa^\mathrm{sca}(\lambda,a)$ and $\kappa^\mathrm{abs}(\lambda,a)$, and the asymmetry parameter $g(\lambda,a)$ determining the scattering phase function $\Phi_{\lambda,a}(\mathbf{k},\mathbf{k'})$, for a range of wavelengths $\lambda$ and a range of grain sizes $a$. Calorimetric properties include the heat capacity $C(T)$ or equivalently the internal energy $U(T)$ of the dust grain material at a range of temperatures $T$, and the bulk mass density $\rho_\mathrm{bulk}$ of the material. Several sets of standard material properties and often-used size distributions are built-in to SKIRT, as illustrated in Table\,\ref{table:components}(c), and new choices can be easily added.

Configurable dust mixes allow users to experiment with new dust models, or to specify a different spatial distribution for a particular dust population (by splitting it off into a separate dust mix corresponding to a dust component with its own geometry).

\subsection{Dust grids}
\label{sec:dustgrids}

\begin{figure}
  \centering
  \includegraphics[width=0.9\columnwidth,height=0.5\columnwidth]{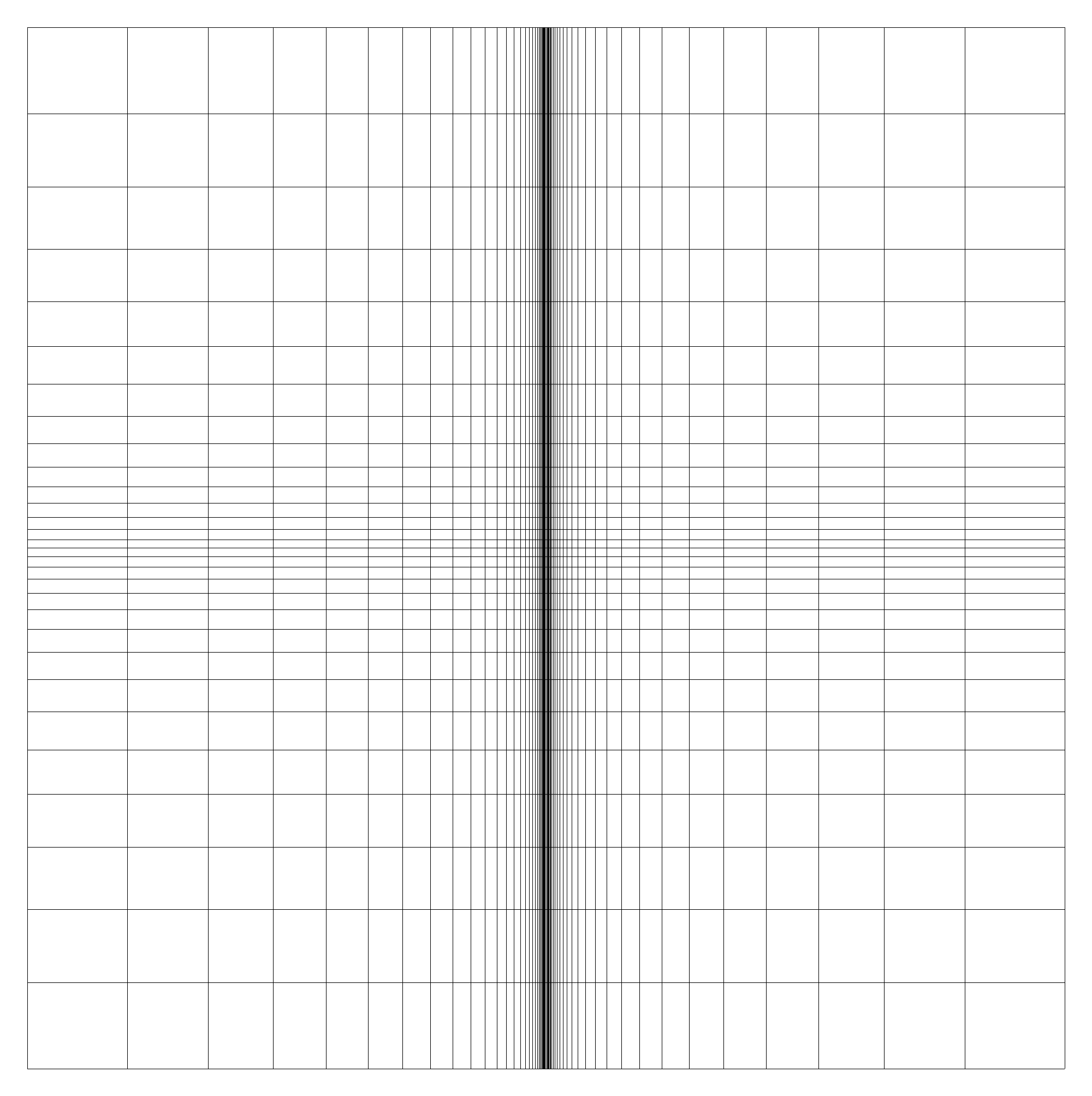}
  \includegraphics[width=0.9\columnwidth]{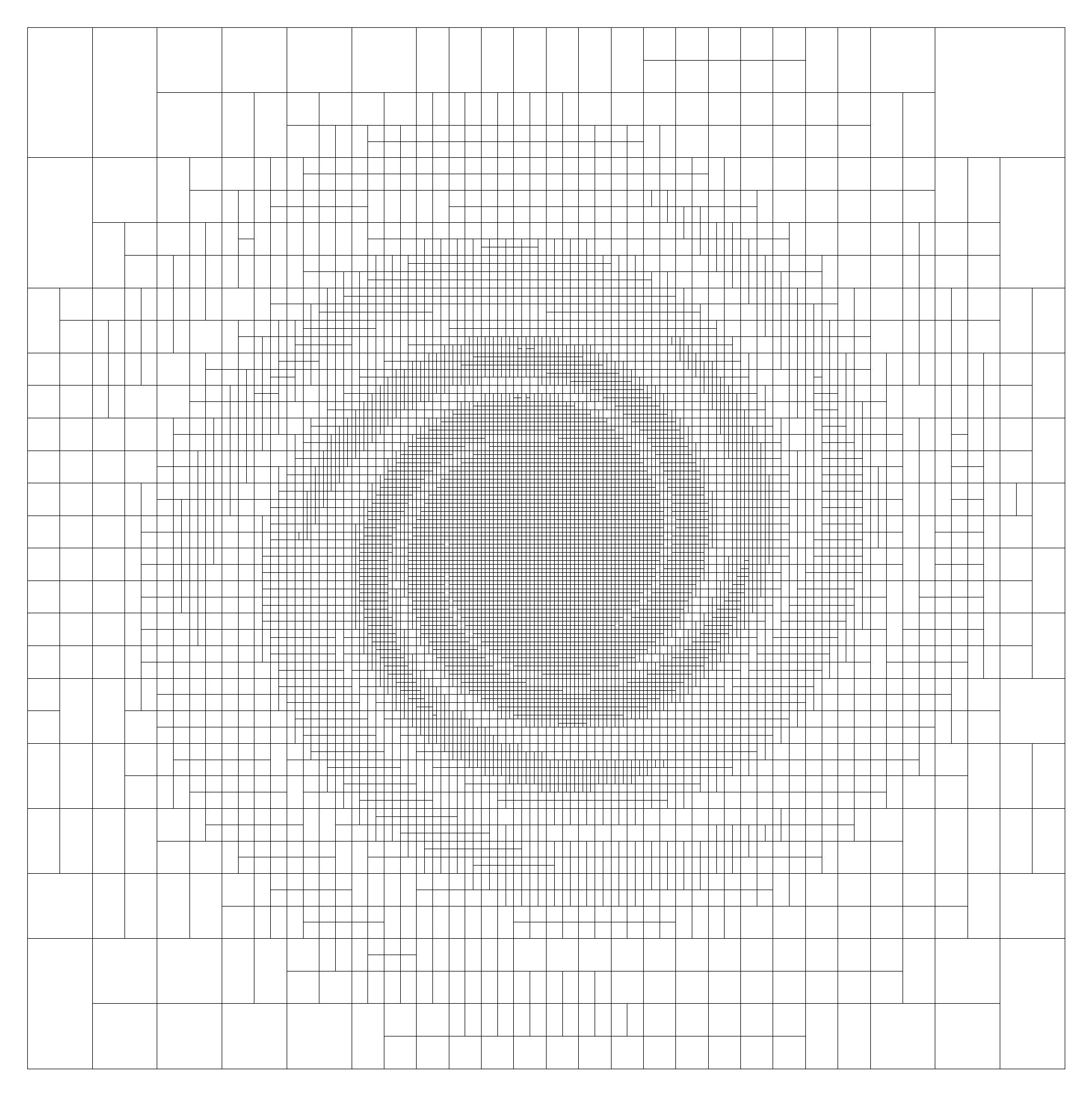}
  \includegraphics[width=0.9\columnwidth]{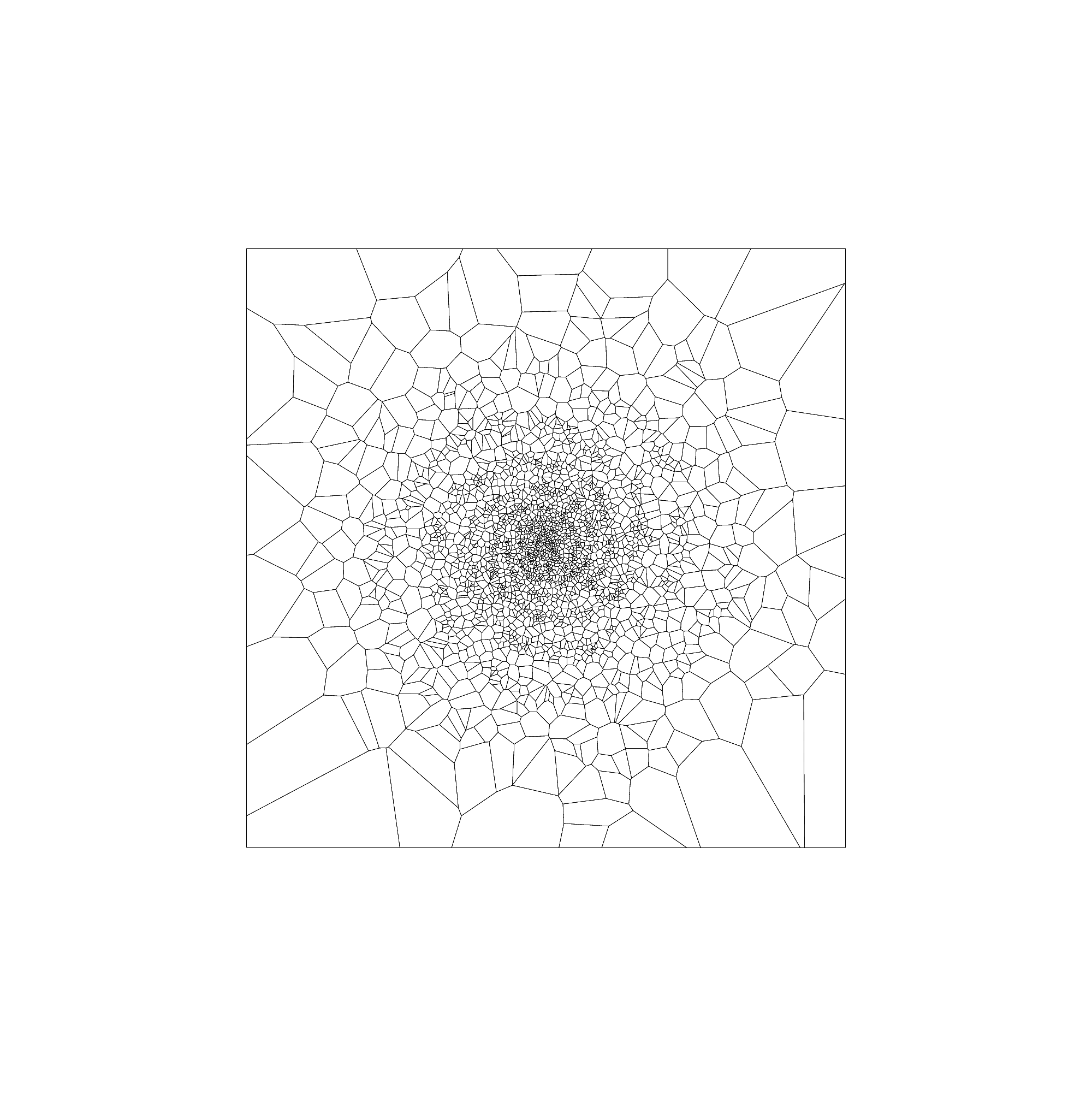}
  \caption{A planar cut through three of SKIRT's dust grids. \emph{Top}: a fixed grid with cell sizes distributed logarithmically on the horizontal axis and according to a power-law on the vertical axis. \emph{Center}: a cuboidal $k$-d tree grid with cell sizes that are adjusted to the dust density distribution in a simple spiral galaxy model. \emph{Bottom}: an unstructured Voronoi grid where the generating sites are placed randomly following the dust density distribution of the same spiral galaxy model. Note that a planar cut through a 3D Voronoi tessellation is usually \emph{not} a 2D Voronoi tessellation.}
  \label{fig:gridcuts}
\end{figure}

The construction of a proper dust grid is a key aspect of a radiative transfer simulation. Many astrophysical models feature small structures, such as dust clumps or star forming regions, which require a lot of cells to resolve properly. To minimize memory requirements and computation time, the grid should be adapted to the spatial structure of the model. Therefore, in addition to fixed grids similar to the one shown in the top panel of Fig.\,\ref{fig:gridcuts}, SKIRT offers several types of adaptive grids, including the $k$-d tree and the Voronoi grid shown in middle and bottom panels of the same figure. 

Starting from a cuboidal root cell that spans the complete spatial domain, a typical adaptive mesh refinement (AMR) scheme recursively subdivides each cell into $a \times b \times c$ cuboidal subcells until sufficient resolution has been reached in each region. In the special case where $a=b=c=2$, each cell is subdivided into eight subcells and the data structure is called an octree. The octree implementation in SKIRT was optimized in the context of radiative transfer as reported in \citet{2013AA...554A..10S}.

A $k$-d tree ($k$-dimensional tree) is a space-partitioning data structure where each cell is recursively split into just two subcells along a particular hyperplane. In 3D space, i.e.\ with $k=3$, a $k$-d tree is similar to an octree. In fact, any octree of depth $n$ has an equivalent $k$-d tree of depth $3n$. For each octree level, the $k$-d tree uses three consecutive levels with mutually orthogonal dividing planes. However, while an octree forces all eight subcells to be created at the same time, a $k$-d tree allows more fine-grained control over which of the two initial subcells are subdivided further. As reported in \citet{2014AA...561A..77S}, this property gives $k$-d tree grids a relevant advantage over octree grids in the context of radiative transfer. The central panel of Fig.\,\ref{fig:gridcuts} shows a cut through a $k$-d tree grid with cell sizes that are adjusted to the dust density distribution in a simple spiral galaxy model.

Adaptive grids with cuboidal cells have become popular mainly because of their relative ease of implementation. But there is no a priori reason to assume that the cuboidal cell form is optimal. To the contrary, the strict coordinate-plane alignment of cell boundaries makes it hard to represent steep gradients in arbitrary directions, raising the number of cells needed to properly resolve clumpy features. One could consider constructing a grid using polyhedra instead of cuboids, but in general this seems a daunting task. Fortunately George Voronoi \citep{crll.1908.134.198} provided a specific way of partitioning 3D space into convex polyhedra. The mathematical properties of a Voronoi tesselation greatly facilitate implementation of an unstructured Voronoi grid in the context of radiative transfer, as described in \citet{2013AA...560A..35C}. Further research should determine whether Voronoi grids can indeed resolve astrophysical structures using less cells than cubodial adaptive grids. The bottom panel of Fig.\,\ref{fig:gridcuts} shows a cut through a 3D Voronoi grid. Cells are placed randomly according to the dust density distribution of a simple spiral galaxy model.

\subsection{User interface}
\label{sec:ui}

The complete configuration for a particular SKIRT simulation is stored in a single parameter file, called a \emph{ski file} (pronounced "skee file"). In view of the many features, options and interdependencies described in the previous sections, the contents of a ski file can become quite complex. To deal with this complexity, we opted for a file format based on XML (eXtensible Markup Language). This format has several advantages. XML elements can be nested to create hierarchies of features and options that reflect the natural makeup of the simulation's configuration. XML is stored as plain text, so it can be easily viewed and adjusted in a regular text editor, even by an occasional user; the human-readable XML tags make the format self-explanatory to a large degree. And finally, existing ski files remain compatible when new features are added (with appropriate defaults), or can be automatically upgraded when the structure changes in an incompatible way.

To perform a simulation, the user starts the code in a terminal window, supplying the name of the relevant ski file on the command line. The code runs fully unattended and all results are written to output files. A small number of command line options allow overriding some defaults in the run-time environment, such as the number of parallel threads or the location of input and output files. The makeup of the simulation itself is fully defined in the ski file. When SKIRT is started without any command line arguments, it enters an interactive query and answer mode that guides the user through the process of creating a new ski file. 

\begin{figure}
  \centering
  \includegraphics[width=0.99\columnwidth]{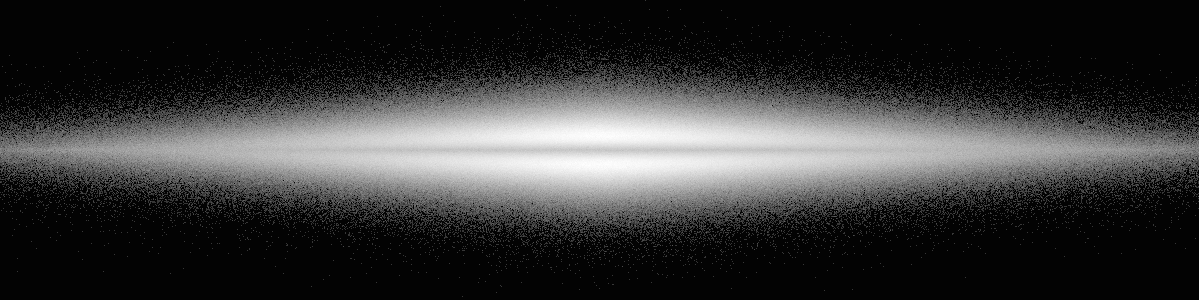}
  \caption{The output of a SKIRT simulation for a very simple spiral galaxy model with an instrument that registers the total flux of an edge-on view. The SKIRT parameter file for this simulation is shown in Fig.\,\ref{fig:spiralski}.}
  \label{fig:spiralflux}
\end{figure}

\begin{figure*}
\centering
\begin{minipage}[c]{0.8\textwidth}
\begin{lstlisting}[style=qasession]
$ skirt
? Enter the name of the ski file to be created: spiralgalaxy
  Possible choices for the simulation:
     1. An oligochromatic Monte Carlo simulation
     2. A panchromatic Monte Carlo simulation
? Enter one of these numbers [1,2] (1): 1
  Possible choices for the units system:
     1. SI units
     2. Stellar units (length in AU, distance in pc)
     3. Extragalactic units (length in pc, distance in Mpc)
? Enter one of these numbers [1,3] (3):
? Enter the number of photon packages per wavelength [0,2e13] (1e6): 1e7
  Possible choices for the wavelength grid:
     1. A list of one or more distinct wavelengths
  Automatically selected the only choice: 1
? Enter the wavelengths [0.0001 micron,1e6 micron]: 0.55
  Possible choices for the stellar system:
     1. A stellar system composed of various stellar components
     2. A stellar system derived from an SPH output file
     ...
? Enter one of these numbers [1,4] (1): 1
     ...
  Possible choices for the geometry of the dust component:
     1. A point source geometry
     2. A Plummer geometry
     ...
     9. An exponential disk geometry
     ...
? Enter one of these numbers [1,26] (9): 9
? Enter the radial scale length [0 pc,*$\infty$* pc]: 6600
? Enter the axial scale height [0 pc,*$\infty$* pc]: 250
? Enter the radial truncation length (zero means no truncation) [0 pc,*$\infty$* pc] (0 pc):
? Enter the axial truncation height (zero means no truncation) [0 pc,*$\infty$* pc] (0 pc):
  Possible choices for the dust mixture of the dust component:
     1. A Draine & Li (2007) dust mix
     ...
? Enter one of these numbers [1,11] (2): 1
  Possible choices for the type of normalization for the dust component:
     1. Normalization by defining the total dust mass
     2. Normalization by defining the edge-on optical depth at some wavelength
     ...
? Enter one of these numbers [1,7] (1): 1
? Enter the total dust mass of the dust component [0 Msun,*$\infty$* Msun]: 4e7
  Possible choices for item #2 in the dust components list:
     1. A dust component
? Enter one of these numbers or zero to terminate the list [0,1] (1): 0
  Possible choices for the dust grid structure:
     1. A cylindrical grid structure with a linear distribution
     ...
     4. A cylindrical grid structure with a radial logarithmic and axial power-law distribution
     ...
    13. A Voronoi dust grid structure
? Enter one of these numbers [1,13] (10): 4
? Enter the inner radius in the radial direction [0 pc,*$\infty$* pc]: 250
? Enter the outer radius in the radial direction [0 pc,*$\infty$* pc]: 25000
? Enter the number of radial grid points [5,100000] (250): 101
? Enter the outer radius in the axial direction [0 pc,*$\infty$* pc]: 7000
? Enter the ratio of the inner- and outermost bin widths in the axial direction [0,1e4] (50): 50
? Enter the number of axial grid points [5,100000] (250): 101
  Successfully created ski file: spiralgalaxy.ski
$
\end{lstlisting}
\end{minipage}
\caption{A partial transcript of the query and answer terminal session to configure a SKIRT simulation for a simple spiral galaxy model. The smart mechanism guides the user through all possible options, narrowing down the possibilities based on earlier choices. For example, on line 15 there is only one choice for the wavelength grid because on line 6 the user selected an oligochromatic simulation. Also the dust grid choices on lines 47-52 are limited to 2D and 3D grids (omitting 1D grids) since the geometry selected on line 29 is axisymmetric. Furthermore the options for the geometry in lines 30-33 and for the dust grid in lines 54-59 are tailored to the selected type of geometry/dust grid.}
\label{fig:spiralterm}
\end{figure*}

\begin{figure*}
\centering
\begin{minipage}[c]{0.8\textwidth}
\begin{lstlisting}[style=skifile]
<?xml version="1.0" encoding="UTF-8"?>
<skirt-simulation-hierarchy type="MonteCarloSimulation" format="6.1">
    <OligoMonteCarloSimulation packages="1e7">
        <units type="Units">
            <ExtragalacticUnits/>
        </units>
        <wavelengthGrid type="OligoWavelengthGrid">
            <OligoWavelengthGrid wavelengths="0.55 micron"/>
        </wavelengthGrid>
        <stellarSystem type="StellarSystem">
            <CompStellarSystem>
                <components type="StellarComp">
                    <OligoStellarComp luminosities="1e11">
                        <geometry type="Geometry">
                            <ExpDiskGeometry radialScale="4400 pc" axialScale="500 pc"
                                radialTrunc="0 pc" axialTrunc="0 pc"/>
                        </geometry>
                    </OligoStellarComp>
                </components>
            </CompStellarSystem>
        </stellarSystem>
        <dustSystem type="OligoDustSystem">
            <OligoDustSystem>
                <dustDistribution type="DustDistribution">
                    <CompDustDistribution>
                        <components type="DustComp">
                            <DustComp>
                                <geometry type="Geometry">
                                    <ExpDiskGeometry radialScale="6600 pc" axialScale="250 pc"
                                        radialTrunc="0 pc" axialTrunc="0 pc"/>
                                </geometry>
                                <mix type="DustMix">
                                    <DraineLiDustMix/>
                                </mix>
                                <normalization type="DustCompNormalization">
                                    <DustMassDustCompNormalization dustMass="4e7 Msun"/>
                                </normalization>
                            </DustComp>
                        </components>
                    </CompDustDistribution>
                </dustDistribution>
                <dustGridStructure type="DustGridStructure">
                    <LogPowAxDustGridStructure
                        radialInnerExtent="250 pc" radialOuterExtent="25000 pc" radialPoints="101"
                        axialExtent="7000 pc" axialRatio="50" axialPoints="101"/>
                </dustGridStructure>
            </OligoDustSystem>
        </dustSystem>
        <instrumentSystem type="InstrumentSystem">
            <InstrumentSystem>
                <instruments type="Instrument">
                    <FrameInstrument instrumentName="xz" distance="10 Mpc"
                        inclination="90 deg" azimuth="-90 deg" positionAngle="0 deg"
                        pixelsX="1200" extentX="28000 pc" pixelsY="300" extentY="7000 pc"/>
                </instruments>
            </InstrumentSystem>
        </instrumentSystem>
    </OligoMonteCarloSimulation>
</skirt-simulation-hierarchy>
\end{lstlisting}
\end{minipage}
\caption{The ski file (SKIRT parameter file) configured during the query and answer session shown in Fig.\,\ref{fig:spiralterm}. While it would be hard for a human to create this file from scratch, it is surprisingly readable because of the self-explanatory tag names. For example, it is easy even for a casual user to adjust the scale height of the dust lane on line 29 or to add an extra instrument by copying lines 52-54 and modifying the inclination angle of the second instrument.}
\label{fig:spiralski}
\end{figure*}

\begin{figure*}
\centering
\begin{minipage}[c]{0.8\textwidth}
\begin{lstlisting}[style=skitex]
*\textbf{SKIRT parameter overview: spiralgalaxy}*

*An oligochromatic Monte Carlo simulation*
*.\quad{}The random number generator: the default random generator*
*.\quad{}.\quad{}The seed for the random generator: $4357$*
*.\quad{}The units system: extragalactic units (length in pc, distance in Mpc)*
*.\quad{}The instrument system: an instrument system*
*.\quad{}.\quad{}Item \#1 in the instruments list: a basic instrument that outputs the total flux in every pixel as a data cube*
*.\quad{}.\quad{}.\quad{}The name for this instrument: xz*
*.\quad{}.\quad{}.\quad{}The distance to the system: $10\:\textrm{Mpc}$*
*.\quad{}.\quad{}.\quad{}The inclination angle $\theta$ of the detector: $90\:^{\circ}$*
*.\quad{}.\quad{}.\quad{}The azimuth angle $\varphi$ of the detector: $\textrm{-}90\:^{\circ}$*
*.\quad{}.\quad{}.\quad{}The position angle $\omega$ of the detector: $0\:^{\circ}$*
*.\quad{}.\quad{}.\quad{}The number of pixels in the horizontal direction: $1200$*
*.\quad{}.\quad{}.\quad{}The maximal horizontal extent: $28000\:\textrm{pc}$*
*.\quad{}.\quad{}.\quad{}The number of pixels in the vertical direction: $300$*
*.\quad{}.\quad{}.\quad{}The maximal vertical extent: $7000\:\textrm{pc}$*
*.\quad{}The number of photon ($\gamma$) packages per wavelength: $1\times 10^{7}$*
*.\quad{}The wavelength grid: a list of one or more distinct wavelengths*
*.\quad{}.\quad{}The wavelengths ($\lambda$): $0.55\:\mu\textrm{m}$*
*.\quad{}The stellar system: a stellar system composed of various stellar components*
*.\quad{}.\quad{}Item \#1 in the stellar components list: a stellar component in an oligochromatic simulation*
*.\quad{}.\quad{}.\quad{}The geometry of the stellar distribution: an exponential disk geometry*
*.\quad{}.\quad{}.\quad{}.\quad{}The radial scale length: $4400\:\textrm{pc}$*
*.\quad{}.\quad{}.\quad{}.\quad{}The axial scale height: $500\:\textrm{pc}$*
*.\quad{}.\quad{}.\quad{}.\quad{}The radial truncation length (zero means no truncation): $0\:\textrm{pc}$*
*.\quad{}.\quad{}.\quad{}.\quad{}The axial truncation height (zero means no truncation): $0\:\textrm{pc}$*
*.\quad{}.\quad{}.\quad{}The luminosities, one for each wavelength, in solar units: $1\times 10^{11}$*
*.\quad{}The dust system: a dust system for use with oligochromatic simulations*
*.\quad{}.\quad{}The dust distribution: a dust distribution composed of various dust components*
*.\quad{}.\quad{}.\quad{}Item \#1 in the dust components list: a dust component*
*.\quad{}.\quad{}.\quad{}.\quad{}The geometry of the dust component: an exponential disk geometry*
*.\quad{}.\quad{}.\quad{}.\quad{}.\quad{}The radial scale length: $6600\:\textrm{pc}$*
*.\quad{}.\quad{}.\quad{}.\quad{}.\quad{}The axial scale height: $250\:\textrm{pc}$*
*.\quad{}.\quad{}.\quad{}.\quad{}.\quad{}The radial truncation length (zero means no truncation): $0\:\textrm{pc}$*
*.\quad{}.\quad{}.\quad{}.\quad{}.\quad{}The axial truncation height (zero means no truncation): $0\:\textrm{pc}$*
*.\quad{}.\quad{}.\quad{}.\quad{}The dust mixture of the dust component: a Draine \& Li (2007) dust mix*
*.\quad{}.\quad{}.\quad{}.\quad{}.\quad{}Output a data file with the optical properties of the dust mix: yes*
*.\quad{}.\quad{}.\quad{}.\quad{}.\quad{}Output a data file with the mean optical properties of the dust mix: yes*
*.\quad{}.\quad{}.\quad{}.\quad{}The type of normalization for the dust component: normalization by defining the total dust mass*
*.\quad{}.\quad{}.\quad{}.\quad{}.\quad{}The total dust mass of the dust component: $4\times 10^{7}\:\textrm{M}_{\odot}$*
*.\quad{}.\quad{}The dust grid structure: a cylindrical grid structure with a radial logarithmic and axial power-law distribution*
*.\quad{}.\quad{}.\quad{}Output data files for plotting the structure of the grid: yes*
*.\quad{}.\quad{}.\quad{}The inner radius in the radial direction: $250\:\textrm{pc}$*
*.\quad{}.\quad{}.\quad{}The outer radius in the radial direction: $25000\:\textrm{pc}$*
*.\quad{}.\quad{}.\quad{}The number of radial grid points: $101$*
*.\quad{}.\quad{}.\quad{}The outer radius in the axial direction: $7000\:\textrm{pc}$*
*.\quad{}.\quad{}.\quad{}The ratio of the inner- and outermost bin widths in the axial direction: $50$*
*.\quad{}.\quad{}.\quad{}The number of axial grid points: $101$*
*.\quad{}.\quad{}The number of random density samples for determining cell mass: $100$*
*.\quad{}.\quad{}Output a data file with convergence checks on the dust system: yes*
*.\quad{}.\quad{}Output FITS files displaying the dust density distribution: yes*
*.\quad{}.\quad{}Calculate and output quality metrics for the dust grid: no*
*.\quad{}.\quad{}Output a data file with relevant properties for all dust cells: no*
*.\quad{}.\quad{}Output statistics on the number of cells crossed per path: no*
\end{lstlisting}
\end{minipage}
\caption{A pretty-printed version of the ski file (SKIRT parameter file) shown in Fig.\,\ref{fig:spiralski}. SKIRT produces a \LaTeX\ file describing the configuration in this way for each simulation performed. The description includes any default values that were omitted from the ski file; see for example lines 50-55.}
\label{fig:spiraltex}
\end{figure*}

To illustrate how this works, we configure a SKIRT simulation for a simple spiral galaxy model, with an instrument that produces the edge-on view shown in Fig.\,\ref{fig:spiralflux}. The query and answer session in the terminal window is illustrated in Fig.\,\ref{fig:spiralterm}. The resulting ski file is shown in Fig.\,\ref{fig:spiralski}, and a pretty-printed version is shown in Fig.\,\ref{fig:spiraltex}.

When configuring a particular type of simulation for the first time, the query and answer mechanism guides the user through all possible options, narrowing down the possibilities based on earlier choices. This is similar to the concept of a \emph{wizard} in graphical user interfaces. Subsequently, the user can easily adjust the constructed ski file in a text editor; a slightly more experienced user can copy and paste building blocks between different ski files. For each simulation performed, SKIRT produces a \LaTeX\ file describing the contents of the input ski file in a human-readable format that can be used for documentation purposes.

To further facilitate the configuration process, physical quantities such as distances, sizes or masses can be specified in units selected by the user. The default unit system for a simulation's input and output is specified early on in the ski file (e.g.\ extragalactic units on line 5 of Fig.\,\ref{fig:spiralski}), and individual parameter values can be specified with a units string that overrides the default. For example, a scale length of $6600\,\mathrm{pc}$ could be specified as {\ttfamily "6600 pc"}, {\ttfamily "6.6 kpc"}, or approximately {\ttfamily "2e20 m"}.

\section{Architecture}
\label{sec:architecture}

The latest version of SKIRT was re-architected with the following major design goals in mind:
\begin{itemize}
\item \emph{Structured parameter file}: use a self-documenting ski file format that supports the complex configuration needs described above in a user-friendly manner.
\item \emph{Single point of definition}: define all information relating to a new feature only once and in the same place, including the code, the human-readable text strings used in the query and answer session, and the tags in the ski file.
\item \emph{Data-driven user interface}: conduct the query and answer session and handle the ski file based solely on these data definitions, so that the user interface adjusts automatically as new features are added.
\item \emph{Shared-memory parallelization}: make all code reentrant by eliminating the use of global variables; protect the remaining global resources or writable shared data with appropriate locking mechanisms.
\item \emph{Modularity}: minimize dependencies among different areas of the code by providing appropriate interfaces and data encapsulation.
\end{itemize}

In this section we describe the overall architecture of the code and we point out how it achieves these design goals. SKIRT is written in C++ using object-oriented design principles. Specifically we use several of the design patterns originally described by \citet{design-patterns} in their classic work, including for example the \emph{Composite}, \emph{Builder}, \emph{Visitor}, and \emph{Decorator} patterns.

\subsection{Simulation items}

\begin{figure}
  \centering
  \includegraphics[width=0.99\columnwidth]{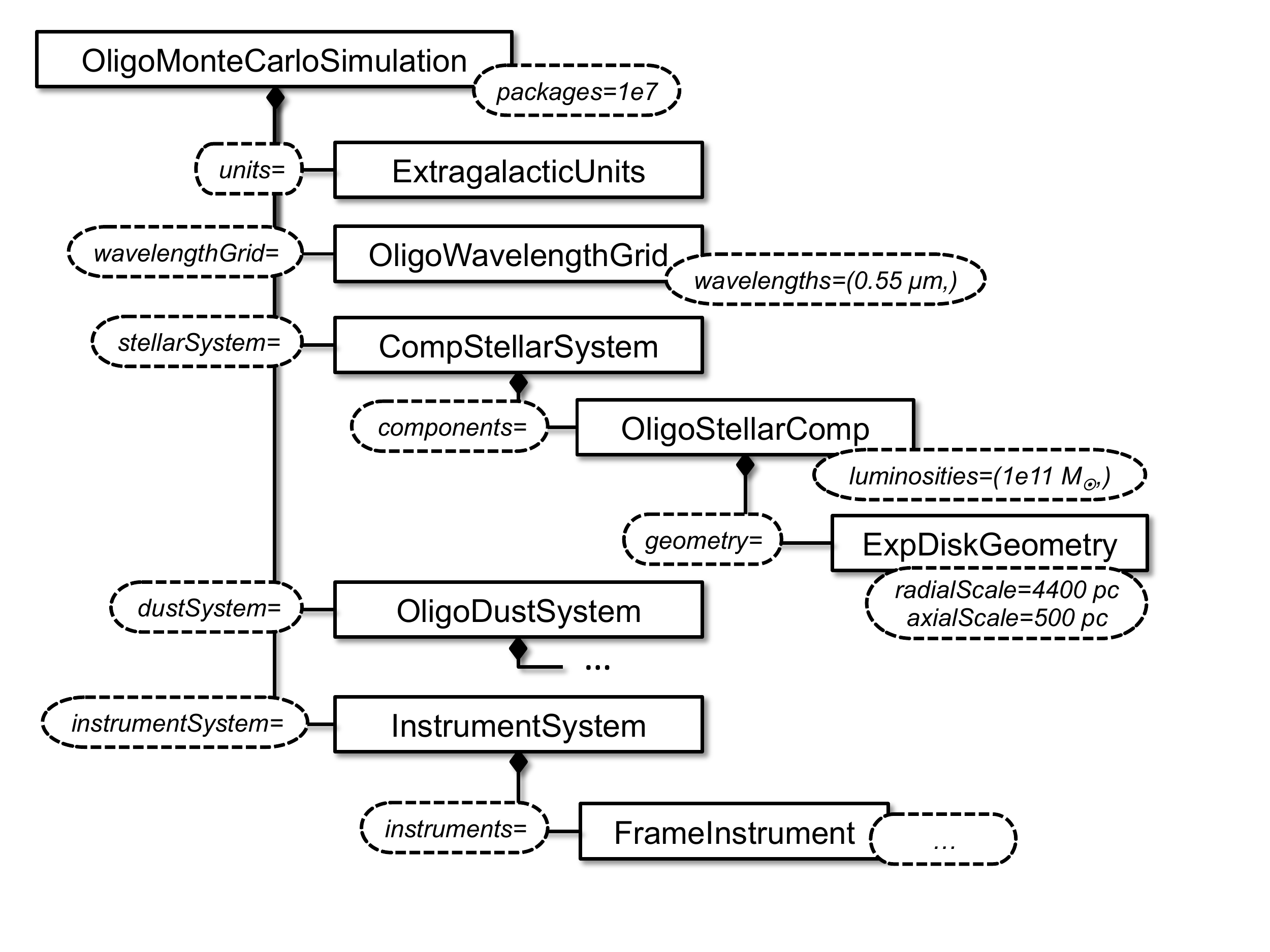}
  \caption{The run-time object hierarchy that would be constructed for the ski file shown in Fig.\,\ref{fig:spiralski}. A solid rectangle represents a simulation item of the specified type; a dashed oval indicates the name and value of a (plain or composite) property. Connections starting with a diamond indicate aggregation. Each simulation item instance and each property in this hierarchy maps directly to an XML element or attribute in the ski file with the same name.}
  \label{fig:runtime-simitems}
\end{figure}

\begin{figure}
  \centering
  \includegraphics[width=0.9\columnwidth]{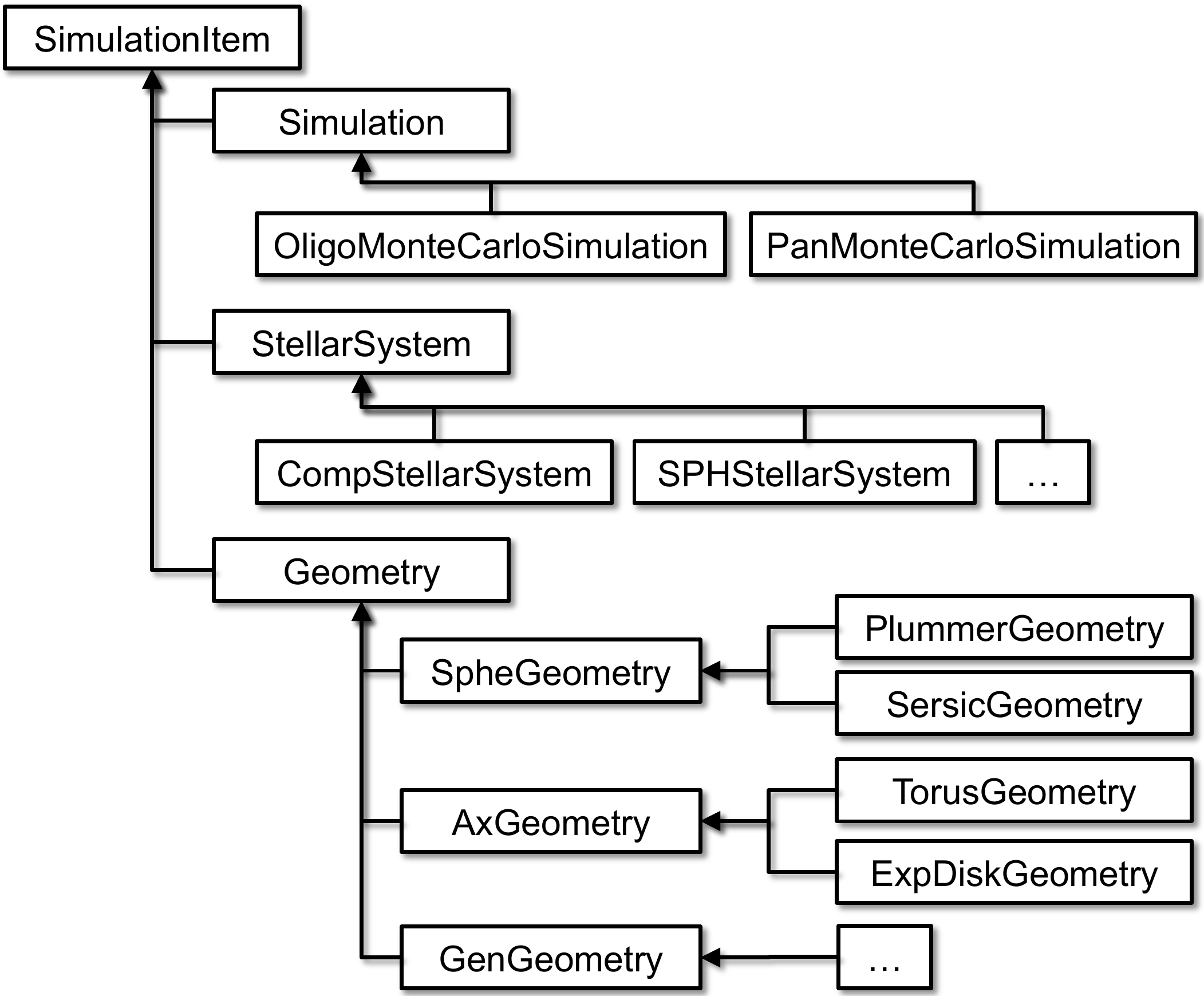}
  \caption{A small portion of the more than 150 simulation item classes in the compile-time inheritance hierarchy. A solid rectangle represents a simulation item class with the specified name. Connections starting with an inverted arrow indicate inheritance.}
  \label{fig:inherit-simitems}
\end{figure}

\begin{figure}
  \centering
  \includegraphics[width=0.6\columnwidth]{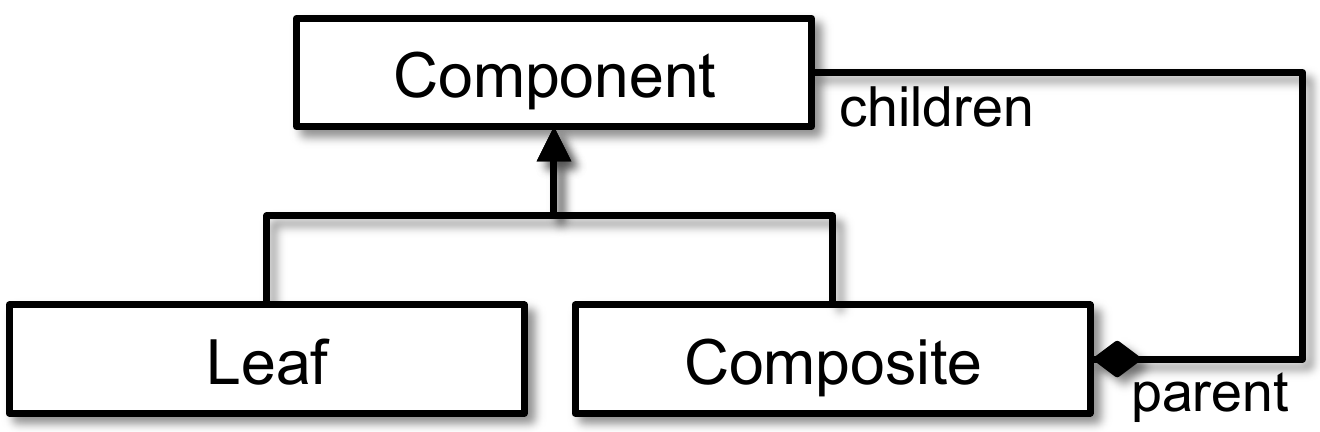}
  \caption{The \emph{Composite} design pattern \citep{design-patterns}. Connections starting with an inverted arrow indicate inheritance; connections starting with a diamond indicate aggregation. This pattern describes an aggregation of objects that all have the same base type.}
  \label{fig:composite-pattern}
\end{figure}

\begin{figure}
  \centering
  \includegraphics[width=0.6\columnwidth]{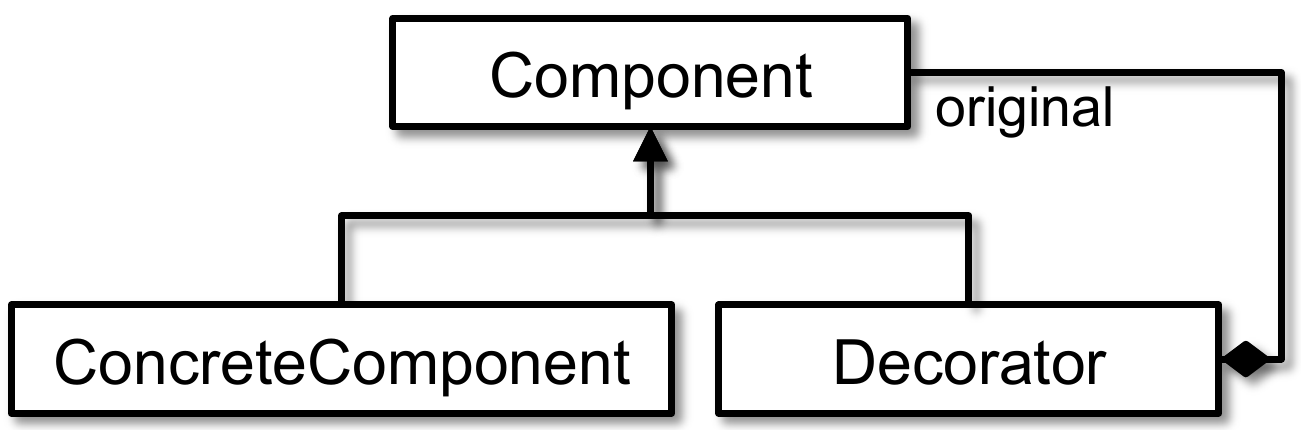}
  \caption{The \emph{Decorator} design pattern \citep{design-patterns}. Connections starting with an inverted arrow indicate inheritance; connections starting with a diamond indicate aggregation. This pattern describes a convenient way to adjust or \emph{decorate} the behavior of another object of the same base type.}
  \label{fig:decorator-pattern}
\end{figure}

\begin{figure}
  \centering
  \includegraphics[width=0.95\columnwidth]{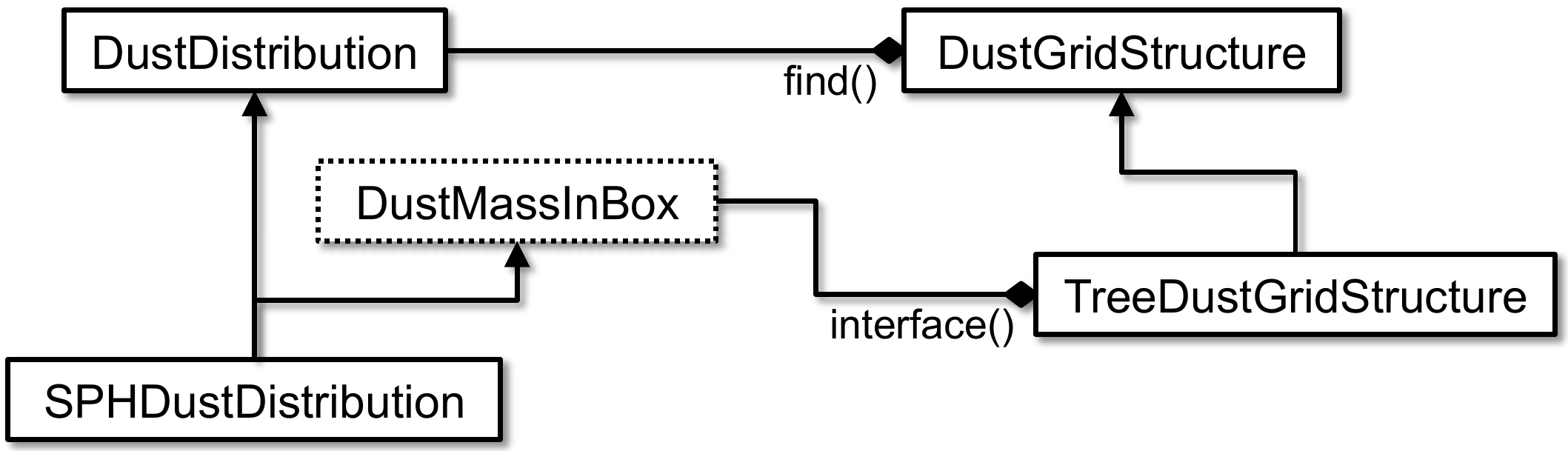}
  \caption{A specialty interface (dotted rectangle) connects two specific simulation items at run time, optimizing performance without creating undesirable dependencies in the respective base classes. Connections starting with an inverted arrow indicate inheritance; connections starting with a diamond indicate aggregation.}
  \label{fig:findinterface}
\end{figure}

\begin{figure}
\centering
\begin{minipage}[c]{0.85\columnwidth}
\begin{lstlisting}[style=cpp]
void SimulationItem::setup()
{
    setupSelfBefore();
    for (SimulationItem* child : children())
    {
        child->setup();
    }
    setupSelfAfter();
}
\end{lstlisting}
\end{minipage}
\caption{The implementation of the \cpp{SimulationItem::setup()} function (ignoring some implementation details).}
\label{fig:setupfun}
\end{figure}

The core of the SKIRT code is obviously about performing radiative transfer simulations. A complete SKIRT simulation is represented at run-time as a hierarchy of objects called \emph{simulation items}, similar to the structure illustrated in Fig.\,\ref{fig:configuration}. This object hierarchy represents the configuration of the simulation (in its structural makeup and in some data members), offers the functionality to perform the simulation (through its member functions), and provides space for any intermediate and resulting data structures (in its data members). Multiple simulation object hierarchies can co-exist and are fully independent of each other. 

The object hierarchy for a particular simulation closely mimics the structure of the corresponding ski file. For example, Fig.\,\ref{fig:runtime-simitems} shows the hierarchy that would be constructed for the ski file listed in Fig.\,\ref{fig:spiralski}. A solid rectangle represents a simulation item of the specified type; a dashed oval indicates the name and value of an property. Plain properties hold a single value (or a list of values); composite properties link other simulation items into the hierarchy. The simulation items and attributes in this hierarchy map directly to an XML element or attribute in the ski file with the same name. This correspondence plays an important role in the automation of the user interface, as we will discuss in Sect.\,\ref{sec:discovery}.

Each simulation item is an instance of a class that directly or indirectly derives from the \cpp{SimulationItem} base class. The simulation item classes form a compile-time inheritance hierarchy, a small portion of which is shown in Fig.\,\ref{fig:inherit-simitems}. The run-time object hierarchy representing a simulation is thus an aggregation of objects of the same type, reflecting the \emph{Composite} design pattern \citep{design-patterns} illustrated in Fig.\,\ref{fig:composite-pattern}. The \role{Component} role is played by the \cpp{SimulationItem} class, and the \role{Composite} role is assumed by any \cpp{SimulationItem} subclass that has one or more composite properties.

The use of the \emph{Composite} pattern is fundamental to the implementation of the user interface discussed in Sect.\,\ref{sec:discovery}, and it substantially facilitates reducing dependencies between portions of the code, as described in Sect.\,\ref{sec:phases} and Sect.\,\ref{sec:dependencies}.

\subsection{Simulation phases}
\label{sec:phases}

A SKIRT simulation has three phases: \emph{construction}, \emph{setup} and \emph{run}. In the \emph{construction} phase, the code constructs the simulation item hierarchy corresponding to a particular ski file, initializing the values of all plain and composite properties, as described in Sect.\,\ref{sec:discovery}. This process completes in a fraction of a second because it doesn't do much work. During the \emph{setup} phase, each simulation item in the hierarchy gets a chance to perform further initialization, such as reading data from resource files or pre-computing frequently used information. This phase may require some processing power, for example, to set up a dust grid that is adapted to the specified dust distribution. Finally, the \emph{run} phase performs the actual simulation and writes down the results. Usually this phase consumes the bulk of the computing resources, and it is fully parallelized.

The \cpp{SimulationItem} base class offers the \cpp{setup()} function; its implementation is shown in Fig.\,\ref{fig:setupfun}. The \cpp{children()} function used on line 4 returns a list of all simulation items held by any composite property of the current simulation item. The \cpp{setupSelfBefore()} and \cpp{setupSelfAfter()} functions are declared \cpp{virtual} in the \cpp{SimulationItem} base class and are overridden by subclasses that need to perform initialization during the \emph{setup} phase. They are invoked respectively before and after any children of the simulation item have been set up. 

The implementation of the \cpp{setup()} function follows the \emph{Template Method} design pattern \citep{design-patterns} to delegate the actual initialization work to subclasses. To recursively invoke all simulation items in the hierarchy, it relies on the fact that all simulation item classes derive from the same class, which is ensured by the use of the \emph{Composite} design pattern.

SKIRT requires that the root object of the simulation item hierarchy inherits from the \cpp{Simulation} class. This class offers the \cpp{run()} function, which, not surprisingly, executes the \emph{run} phase. Thus, after constructing the run-time hierarchy, the code simply invokes the \cpp{setup()} and \cpp{run()} functions on the hierarchy's root object to complete all phases of the simulation.

\subsection{Reducing dependencies}
\label{sec:dependencies}

Most simulation item classes are organized in groups with a common purpose, e.g.\ wavelength grids, geometries, or dust mixes. All classes in a particular group inherit from the same base class, i.e.\ \cpp{WavelengthGrid}, \cpp{Geometry}, or \cpp{DustMix}. The base class offers the common interface for all classes in the group towards classes outside of the group. This design principle avoids undesirable dependencies between classes in different groups, enhancing modularity.

Some information about a simulation's configuration is accessed from many different places, and thus must be readily available. To facilitate access to other simulation items in the same hierarchy, the \cpp{SimulationItem} class offers the \cpp{T* find<T>()} template function, where \cpp{T} stands for the name of any class that derives from \cpp{SimulationItem}. This template function searches the object hierarchy in which the receiving simulation item resides for a simulation item of the specified type \cpp{T}, and returns a pointer to the first such object found after dynamically casting it to the requested type. If the hierarchy does not contain an object of the specified type, the function throws an exception. The implementation of the \cpp{find<>()} template function again relies on the fact that all simulation item classes derive from the same class.

For example, every run-time simulation hierarchy includes an instance of a particular \cpp{WavelengthGrid} subclass, such as \cpp{OligoWavelengthGrid} or \cpp{NestedLogWavelengthGrid}. Any simulation item in the hierarchy can call \cpp{find<WavelengthGrid>()} to retrieve a pointer to the common wavelength grid interface; the caller does not know the specific sub-type of the returned object. Also, the caller has no need to know where the returned object resides, so this mechanism replaces application-wide global data (which gets in the way of parallelization) by simulation-wide available data.

While modularity is important, the generic and narrow interfaces between different areas of the code sometimes hide information that can be relevant for optimal cooperation between components. As a first example, the dust mass in each grid cell is usually estimated by probing the dust density distribution in a number of random locations uniformly distributed over the spatial extent of the cell. This generic mechanism works for any cell shape and for any type of density distribution. However, for certain cell shapes combined with certain types of density distribution, it might be orders of magnitude faster to directly calculate the mass in the cell. As a second example, sometimes we would like to build a dust grid based on particular locations (such as SPH particles) defined as part of the input dust density distribution, rather than based on the density distribution itself. However, the generic interface does not offer particle information because the concept is meaningless for most density distributions.

These features can be accomplished without breaking modularity by using a specialty interface that is known only to the specific classes involved, and by providing a mechanism to dynamically connect the two players at run time. This is illustrated in Fig.\,\ref{fig:findinterface} for the first example described above. The \cpp{DustMassInBox} interface declares pure virtual functions offering the relevant special capabilities. The \cpp{SPHDustDistribution} class inherits the interface and actually implements its functions. Finally, the \cpp{TreeDustGridStructure} simulation item recovers a pointer to the specialty interface as follows. First it uses the \cpp{find<>()} template function to retrieve the dust distribution object in the hierarchy; this could in fact happen in the \cpp{DustGridStructure} base class since the returned pointer is of the generic type \cpp{DustDistribution}. Then it invokes the \cpp{interface<>()} template function on the dust distribution object, which is in fact of type \cpp{SPHDustDistribution}, to return a pointer to the \cpp{DustMassInBox} interface implemented by that same object.

For this purpose, the \cpp{SimulationItem} base class provides the \cpp{T* interface<T>()}, where \cpp{T} stands for the name of the specialty interface to be recovered. In the example of Fig.\,\ref{fig:findinterface}, the function can be implemented with a simple dynamic cast. To support more complicated cases, the function also allows a simulation item to delegate the implementation of a specialty interface to a different object.

\subsection{Reusing components}
\label{sec:reuse}

In a Monte Carlo radiative transfer code, the key function of a dust geometry (describing the distribution of the dusty medium) is to retrieve the dust density at a specified location in space. This function is called (quite often) during setup to build an appropriate dust grid and calculate the dust mass in each grid cell. On the other hand, the key function of a stellar geometry (describing the distribution of radiation sources) is to generate a random location in space, drawn from a probability distribution corresponding to the geometry's density distribution. This function is called repeatedly during the simulation to determine the point of emission for a new photon package.

It might seem that the functionalities of the two geometry types are thus rather disjunct, but this is not the case. For example, we want to build a Voronoi dust grid using generating sites placed according to the dust density distribution. In this case, we need the key stellar geometry functionality (generating random points) in the dust geometry. Therefore the recent SKIRT version has unified geometry classes that offer both functions, as listed in Table\,\ref{table:components}(a).

\begin{figure}
  \centering
  \includegraphics[width=0.48\columnwidth]{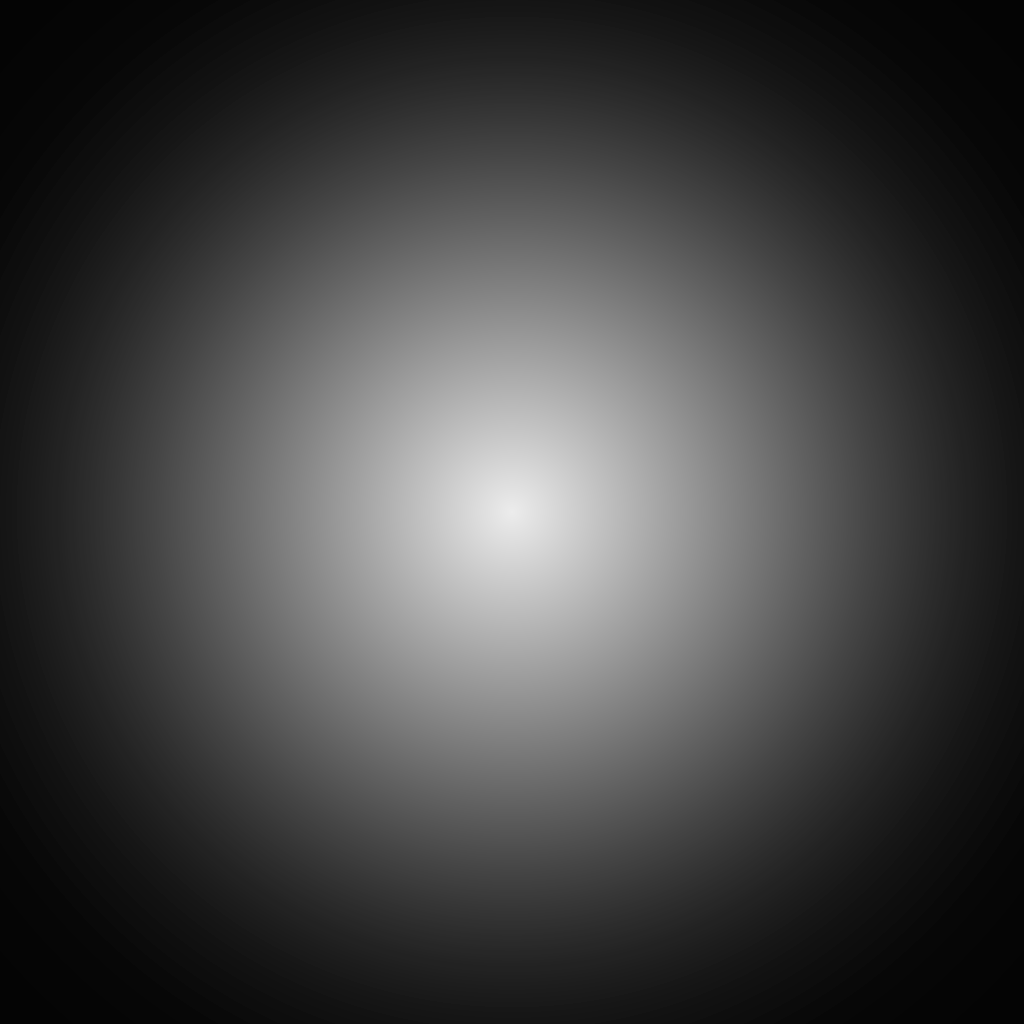}
  \includegraphics[width=0.48\columnwidth]{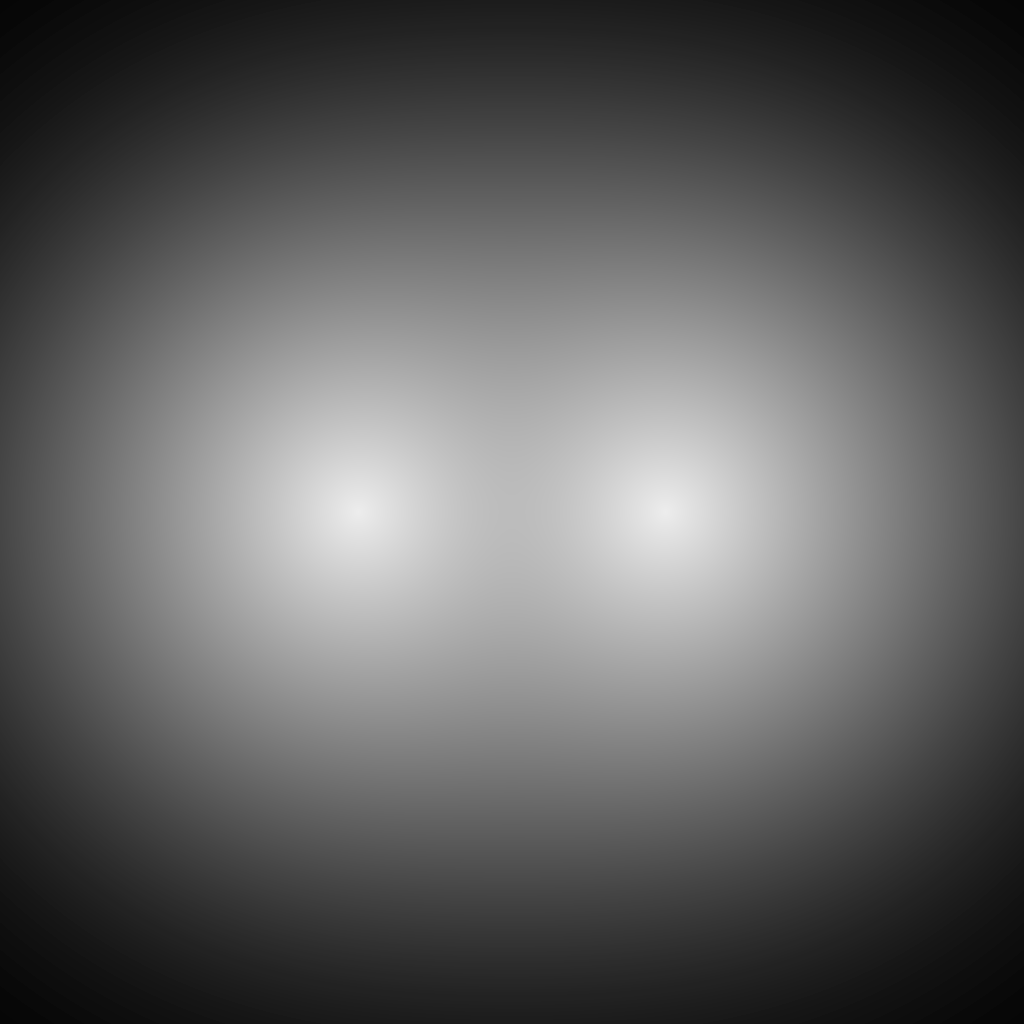}
  \\[2pt]
  \includegraphics[width=0.48\columnwidth]{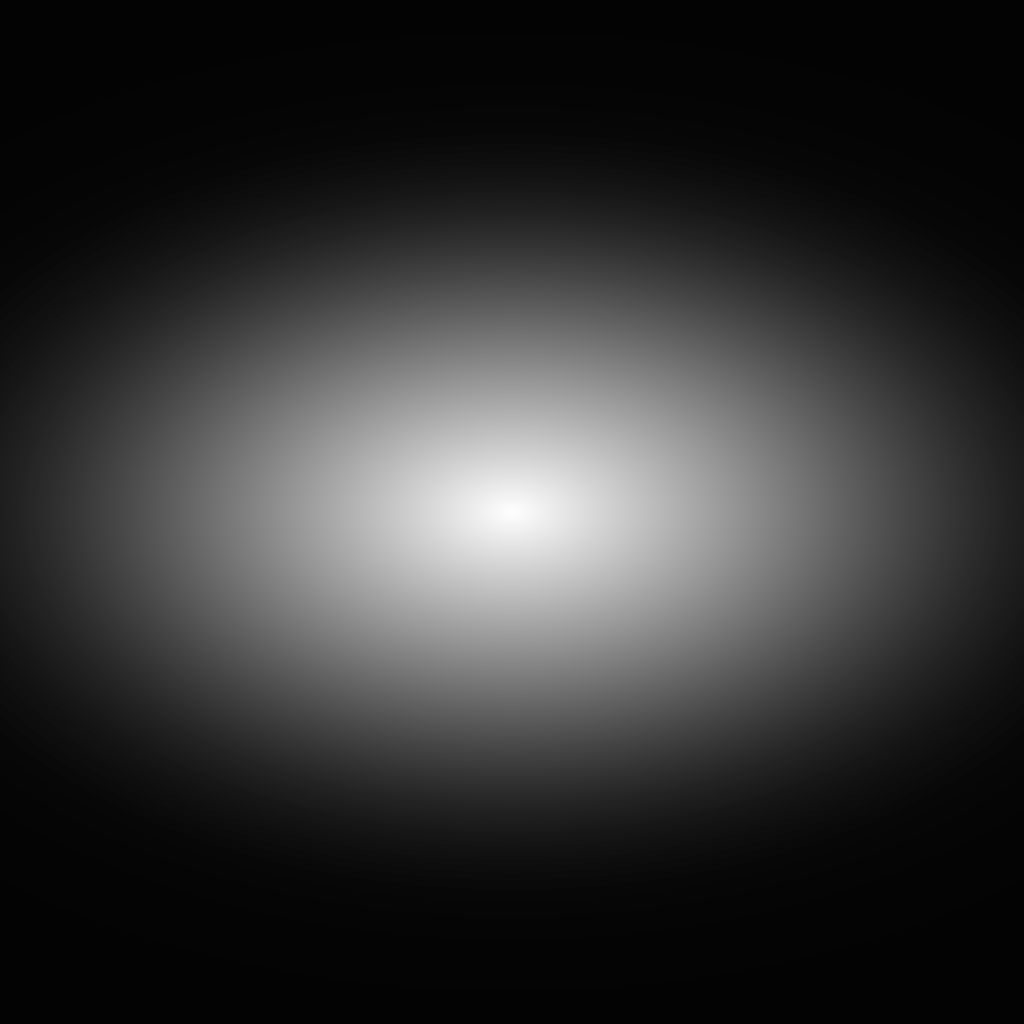}
  \includegraphics[width=0.48\columnwidth]{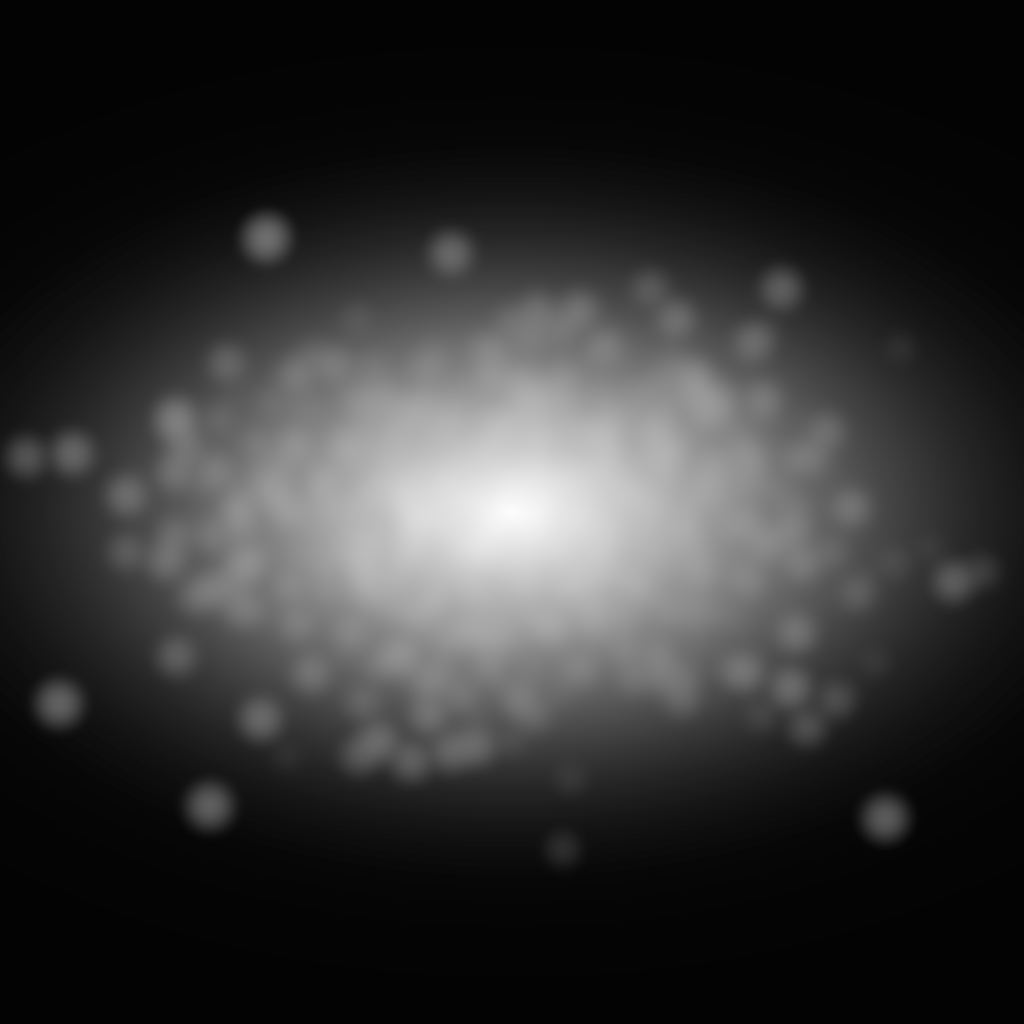}
  \caption{Planar cuts through four density distributions derived from the same underlying geometry using the \emph{Decorator} design pattern: a plain, spherical Einasto profile with index 1 (\emph{top left}); two superposed Einasto profiles, each shifted aside using an offset decorator (\emph{top right}); the Einasto profile deformed into a spheroidal shape by a decorator (\emph{bottom left});  and a clumpy, spheroidal Einasto distribution derived from the original profile by applying a chain of two decorators (\emph{bottom right}).}
  \label{fig:einasto-decorated}
\end{figure}

This change prompted us to invest in a number of geometry classes that modify other geometry's in interesting ways, following the \emph{Decorator} design pattern \citep{design-patterns} shown in Fig.\,\ref{fig:decorator-pattern}. An object in the \role{Decorator} role maintains a pointer to another object of the same base class, called the original object. The decorator implements the base class interface by calling corresponding functions in the original object, and returning the results after possible adjustment. In SKIRT, the \cpp{Geometry} class plays the \role{Component} role in this pattern, and any \cpp{Geometry} subclass can assume the \role{ConcreteComponent} role, i.e.\ the role of the original geometry being decorated. The \cpp{ClumpyGeometry} class is one of the classes that plays the \role{Decorator} role. It modifies the original geometry by replacing a fraction of its total mass allocation by randomly placed clumps. Other decorators relocate the original geometry's center, or deform a spherical geometry into a spheroidal or triaxial distribution. Multiple decorators can be chained to achieve the combined effects, as illustrated in Fig.\,\ref{fig:einasto-decorated}.

\subsection{Automating the user interface}
\label{sec:discovery}

The latest version of SKIRT is based on the Qt development framework\footnote{The Qt project: http://qt-project.org}, which includes a rich set of cross-platform C++ libraries and an integrated development environment (IDE) called Qt Creator. Although we don't need its graphical user interface (GUI) capabilities, the Qt environment offers substantial benefits in other areas as well\footnote{The new features in the recent C++11 language standard cover much of the functionality for which SKIRT uses Qt, with the exception of the introspection capabilities discussed in this section.}. Specifically, the Qt environment provides run-time introspection of classes and their member functions, assuming the appropriate declarations were added in the code. The Qt mechanism is a lot more advanced than the standard C++ run-time type information (RTTI) system. For example, the Qt library offers functions to retrieve compile-time information such as the class inheritance hierarchy and the type of function arguments or return values. It is also possible to invoke a function by specifying the function name at run-time as a string, or to construct a new class instance in a similar manner.

\begin{figure*}
\centering
\begin{minipage}[c]{0.8\textwidth}
\begin{lstlisting}[style=cpp]
class ClumpyGeometry : public Geometry
{
    Q_OBJECT
    Q_CLASSINFO("Title", "a geometry that adds clumpiness to any geometry")

    Q_CLASSINFO("Property", "geometry")
    Q_CLASSINFO("Title", "the geometry to which clumpiness is added")

    Q_CLASSINFO("Property", "clumpFraction")
    Q_CLASSINFO("Title", "the fraction of the mass locked up in clumps")
    Q_CLASSINFO("MinValue", "0")
    Q_CLASSINFO("MaxValue", "1")

    Q_CLASSINFO("Property", "clumpCount")
    Q_CLASSINFO("Title", "the total number of clumps")
    Q_CLASSINFO("MinValue", "1")

    Q_CLASSINFO("Property", "clumpRadius")
    Q_CLASSINFO("Title", "the scale radius of a single clump")
    Q_CLASSINFO("Quantity", "length")
    Q_CLASSINFO("MinValue", "0")

    Q_CLASSINFO("Property", "cutoff")
    Q_CLASSINFO("Title", "cut off clumps at the boundary of the underlying geometry")
    Q_CLASSINFO("Default", "no")

public:
    Q_INVOKABLE ClumpyGeometry();

protected:
    void setupSelfBefore();
    void setupSelfAfter();

public:
    Q_INVOKABLE void setGeometry(Geometry* value);
    Q_INVOKABLE Geometry* geometry() const;

    Q_INVOKABLE void setClumpFraction(double value);
    Q_INVOKABLE double clumpFraction() const;

    Q_INVOKABLE void setClumpCount(int value);
    Q_INVOKABLE int clumpCount() const;

    Q_INVOKABLE void setClumpRadius(double value);
    Q_INVOKABLE double clumpRadius() const;

    Q_INVOKABLE void setCutoff(bool value);
    Q_INVOKABLE bool cutoff() const;

public:
    double density(Position bfr) const;
    Position generatePosition() const;

    ...
};
\end{lstlisting}
\end{minipage}
\caption{A typical simulation item class declaration. The keywords starting with \cpp{Q\_} are provided by the Qt development environment, and serve to define the extra information needed to automatically build a user interface for the features offered by this class, as explained in Sect.\,\ref{sec:discovery}. The setup functions declared on lines 31-32 are described in Sect.\,\ref{sec:phases} and Fig.\,\ref{fig:setupfun}. The geometry-specific functions declared on lines 51-52 are described in Sect.\,\ref{sec:reuse}.}
\label{fig:classinfo}
\end{figure*}

SKIRT relies on the Qt introspection features to automatically construct a user interface from the C++ class declarations in the code. To enable this process, all \cpp{SimulationItem} subclass declarations must be augmented with some extra information, as illustrated in Fig.\,\ref{fig:classinfo}. The keywords starting with \cpp{Q\_} are provided by the Qt development environment. The \cpp{Q\_OBJECT} keyword on line 3 is required to enable the Qt introspection features for this class. The \cpp{Q\_CLASSINFO} definitions on lines 4-25 associate an ordered list of key-value pairs with the compile-time class information; these strings can be retrieved at run-time through the Qt introspection system. The \cpp{Q\_INVOKABLE} keyword on lines 28 and 35-48 enables Qt introspection for the constructor or member function declaration following the keyword. 

In SKIRT, the \cpp{Q\_CLASSINFO} key-value pairs are used to provide a human readable description for the class, and to define its configurable properties (i.e.\ the properties that can be specified in a ski file). The name of a property, e.g.\ \cpp{"clumpFraction"}, must match the name of a getter function \cpp{clumpFraction()} and a setter function \cpp{setClumpFraction()}, declared with the keyword \cpp{Q\_INVOKABLE}. The type of the configurable property is derived from the getter's return value (\cpp{double} in this case). Additional key-value pairs can specify options such as a default value, or the type of physical quantity represented by this property, which determines the units used or accepted in the user interface.

\begin{figure}
  \centering
  \includegraphics[width=0.99\columnwidth]{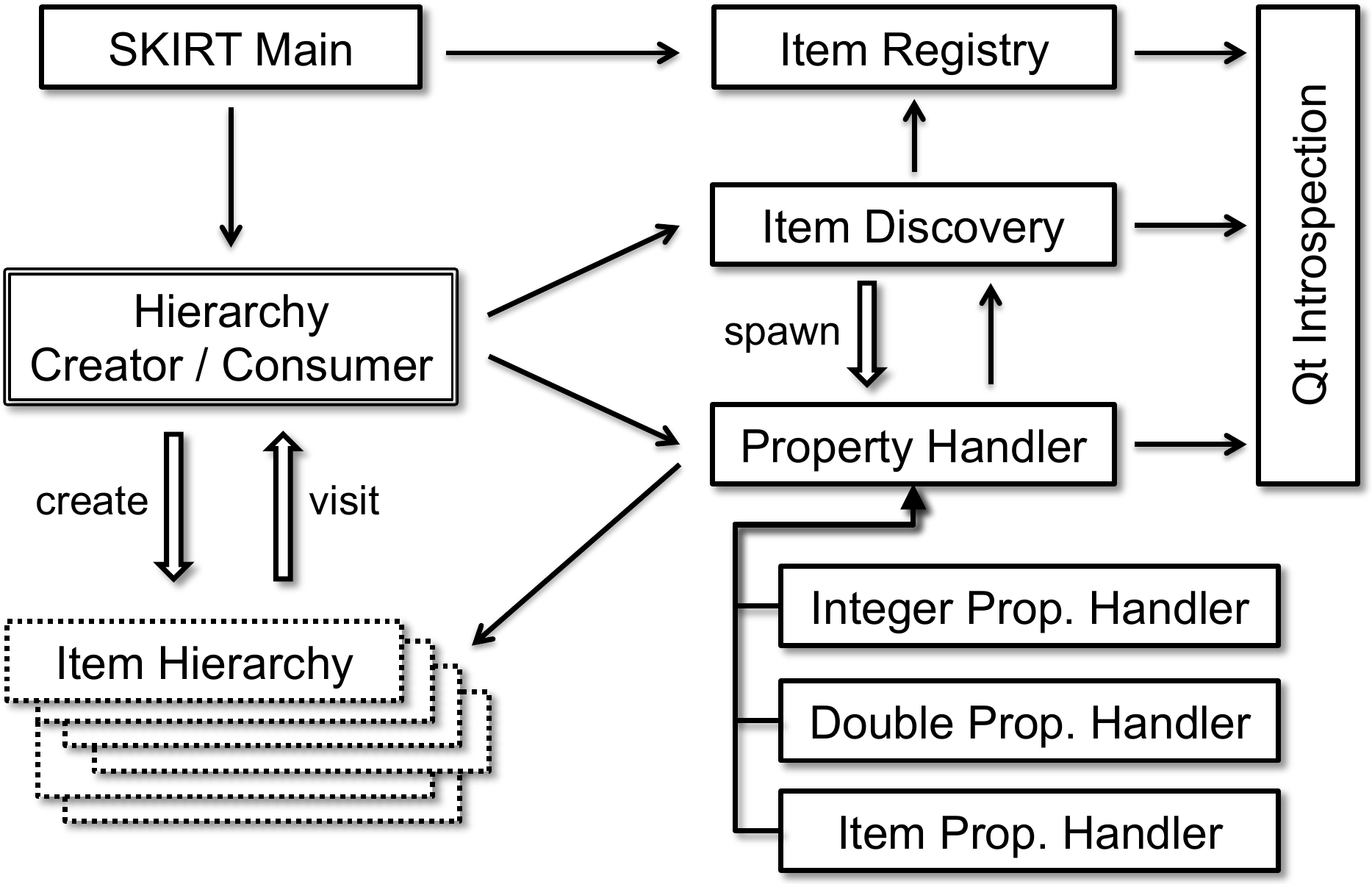}
  \caption{Schematic overview of the code that automatically builds the user interface from simulation item class declarations. In this diagram, the phrase \emph{simulation item} has been shortened to \emph{item}. A solid arrow indicates that the source module uses the target module. Connections starting with an inverted arrow indicate inheritance. The relationships are described in more detail in Sect.\,\ref{sec:discovery}.}
  \label{fig:discovery}
\end{figure}

Figure\,\ref{fig:discovery} depicts the overall organization of the code that automatically builds the user interface from the compile-time data. The architecture is inspired by -- but doesn't correspond exactly to -- the \emph{Builder} and \emph{Visitor} design patterns \citep{design-patterns}. The \emph{hierarchy creator} object in the figure plays the \role{Builder} role, and the \emph{hierarchy consumer} object plays the \role{Visitor} role. 

 Just after program startup, the \emph{item registry} is initialized with a list of all simulation item classes. This ensures that all classes are actually linked into the code, and it provides the starting point for the \emph{item discovery} module to implement queries about the simulation item classes. Functions offered by this module include for example \cpp{title(itemType)}, \cpp{descen\-dants(itemType)}, and \cpp{createPropertyHandlers(item)}. The latter function spawns a property handler of the appropriate type for each property of the specified simulation item.

A property handler combines a pointer to a particular simulation item object in the hierarchy, with knowledge about the compile-time attributes of one of the properties of the item. The handler can be used to get or set the property value directly from or into the target object, or to retrieve attributes such as its default value. The handler also knows how to convert a property value into a string for human consumption, and vice versa. There are handlers for various property types, including Boolean, integer, enumeration, floating point (with support for units), string, and pointer to simulation item.

To create a simulation item hierarchy from a ski file, the command line handler in the \emph{SKIRT main} module enlists an \cpp{XmlHierarchyCreator} object. This object uses the XML tags in the ski file, which correspond to simulation class names and property names, to recursively construct the corresponding simulation items and set their property values. The code heavily relies on the \emph{item discovery} module and the property handlers spawned by it.

Similarly, a \cpp{ConsoleHierarchyCreator} object is used to create a hierarchy from scratch by conducting a query and answer session. The top of the hierarchy must be occupied by an instance of the \cpp{Simulation} class, so the algorithm obtains a list of concrete \cpp{Simulation} subclasses from the \emph{item discovery} module, and asks the user to make a choice. The question is formulated using the titles provided in each class declaration; see lines 4-6 in Fig.\,\ref{fig:spiralterm}. The algorithm constructs an instance of the selected subclass, and then loops over all of its configurable properties, asking the appropriate question(s) for each property depending on its type. Boolean, numeric and string properties only need a single question; see lines 30-33 and 54-59 in Fig.\,\ref{fig:spiralterm}. A property that points to another simulation item prompts a multiple choice question to select one of the available concrete subclasses that inherit the appropriate type, again obtained from the \emph{item discovery} module; see lines 7-11 and 23-29 in Fig.\,\ref{fig:spiralterm}. A new simulation item of the selected type is created (and linked into the hierarchy), and the same mechanism is recursively applied to the new object.

By selecting the desired type of simulation item at each level in the recursion, the user's responses drive the nature of subsequent questions in the session. While this is sufficient for most purposes, the discovery process implements a few extra mechanisms to support specific needs. For example, the list of available dust grids depends on the (lack of) symmetries in the geometries selected earlier; e.g.\ the user can't select a 1D or 2D grid for a 3D geometry. Also, it is possible to skip questions that are deemed irrelevant based on the response to a previous question in the same class. All of these mechanisms are fully data-driven from the \cpp{Q\_CLASSINFO} definitions in the simulation item class declarations.

Once a simulation item hierarchy is in place, the same underlying data can be used to reverse the process and write down the configuration in a human-readable form. In Fig.\,\ref{fig:discovery} the \role{Creator} object is now replaced by a \role{Consumer} object that recursively visits the items in the hierarchy to produce the corresponding output, using the information supplied by the \emph{item discovery} module and the property handlers it spawns. Most importantly, SKIRT uses the \cpp{XmlHierarchyWriter} object to output a ski file (Fig.\,\ref{fig:spiralski}) after the user configured a simulation item hierarchy through a query and answer session (Fig.\,\ref{fig:spiralterm}). A newly generated ski file is also stored with each set of simulation results, as a standard reference, explicitly listing the default values for properties that may have been omitted in the input ski file. Using the \cpp{LatexHierarchyWriter} object, SKIRT also writes a \LaTeX\ source file that documents the configuration in an even more user-friendly format (Fig.\,\ref{fig:spiraltex}). 

Because of this automation, adding a new SKIRT feature is extremely straightforward. For example, to add a new geometry, a developer would copy one of the existing geometry classes, rename the class, adjust the implementation and documentation of the member functions, adjust the \cpp{Q\_CLASSINFO} definitions in the class declaration, and add a single line in the \cpp{RegisterSimulationItems} class to register the new geometry to the discovery system. Except for this trivial registration requirement, all information about the new geometry is in a single place, and the user interface will be automatically adjusted to incorporate it.

\section{Conclusions}
\label{sec:conclusions}

We described the major features of SKIRT, a state-of-the-art Monte Carlo dust radiation transfer simulation code used to study spiral galaxies, accretion disks and other astrophysical systems. In addition to its core capability of tracing the radiation through the dust, SKIRT offers a large number of built-in options for configuring all aspects of the simulation model, including spatial and spectral distributions, dust grain characterizations, simulated detection systems, and discretization.

Providing a proper user interface to support these complex configuration requirements is a nontrivial undertaking. Most SKIRT simulations run for hours or days, often on remote servers, so there is no need for a fancy graphical user interface. Still, the typical user is an expert astrophysicist who interacts with the SKIRT code rather occasionally, and thus benefits greatly from a low-barrier interface. SKIRT addresses this challenge through the combination of a \emph{wizard}-like query and answer session to guide a first-time user through the configuration process, and self-documenting XML-based parameter files that can be easily updated in a text editor.

We further described the overall architecture of the code. Inspired by standard software design principles and patterns, the latest version of SKIRT has a modular implementation that can be easily maintained and expanded. Programming interfaces between components are well defined and narrow. The user interface is automatically constructed from data provided in the C++ class declarations, allowing a single point of definition, and placing the user interface information right next to the code implementing the corresponding feature.

All too often, scientific codes are written without much concern for user interface or for modular software design. This is very unfortunate. Scientists may not need a \emph{graphical} user interface, but, just like every one else, they do benefit from an interaction mechanism that hides the underlying complexity. As we have illustrated in this work, a well-designed non-graphical user interface may be a perfect fit, and can often be developed and maintained with limited resources. Similarly, adhering to proven software design principles pays off, even for small and mid-sized projects. 

The SKIRT source code is publicly available, and it has already been applied to radiative transfer problems in various astrophysical domains. We welcome new applications, and we invite potential users and code contributors to join the SKIRT community.


\section*{Acknowledgements}

This work fits in the CHARM framework (Contemporary physical challenges in Heliospheric and AstRophysical Models), a phase VII Interuniversity Attraction Pole (IAP) program organized by BELSPO, the BELgian federal Science Policy Office. 

SKIRT and FitSKIRT are based on the Qt development framework and use the core Qt libraries for run-time class introspection, parallelization, string and container handling, and more. The code further incorporates the following third-party software libraries: CFITSIO developed by NASA's HEASARC to output FITS files; Voro++ described in \citet{Voro-lib} to help construct Voronoi dust grids; GAlib described in \citet{Wall96} to implement the search mechanism based on genetic algorithms.

The authors hereby thank all SKIRT users for their enthusiastic feedback, including many ideas for improvements and additions to the code.

\section*{References}
\bibliography{skirt7}
\end{document}